\catcode`@=11

\def\singlespace{\normalbaselines}
\def\oneandahalfspace{\baselineskip=1.15\normalbaselineskip plus 1pt
\lineskip=2pt\lineskiplimit=1pt}

\def\np{\vfill\eject}
\def\nl{\hfil\break}

\def\nofirstpagenoten{\nopagenumbers\footline={\ifnum\pageno>1\tenrm
\hss\folio\hss\fi}}
\def\nofirstpagenotwelve{\nopagenumbers\footline={\ifnum\pageno>1\twelverm
\hss\folio\hss\fi}}
\def\leaderfill{\leaders\hbox to 1em{\hss.\hss}\hfill}
\def\ft#1#2{{\textstyle{{#1}\over{#2}}}}
\def\frac#1#2{\leavevmode\kern.1em
\raise.5ex\hbox{\the\scriptfont0 #1}\kern-.1em/\kern-.15em
\lower.25ex\hbox{\the\scriptfont0 #2}}
\def\sfrac#1#2{\leavevmode\kern.1em
\raise.5ex\hbox{\the\scriptscriptfont0 #1}\kern-.1em/\kern-.15em
\lower.25ex\hbox{\the\scriptscriptfont0 #2}}

\parindent=20pt
\def\narrow{\advance\leftskip by 40pt \advance\rightskip by 40pt}

\def\AB{\bigskip
        \centerline{\bf ABSTRACT}\medskip\narrow}
\def\nonarrower{\advance\leftskip by -40pt\advance\rightskip by -40pt}
\def\AE{\bigskip\nonarrower}

\def\boxit#1{\vbox{\hrule\hbox{\vrule\kern3pt
        \vbox{\kern3pt#1\kern3pt}\kern3pt\vrule}\hrule}}

\def\gtorder{\mathrel{\raise.3ex\hbox{$>$}\mkern-14mu
             \lower0.6ex\hbox{$\sim$}}}
\def\ltorder{\mathrel{\raise.3ex\hbox{$<$}|mkern-14mu
             \lower0.6ex\hbox{\sim$}}}
\def\dalemb#1#2{{\vbox{\hrule height .#2pt
        \hbox{\vrule width.#2pt height#1pt \kern#1pt
                \vrule width.#2pt}
        \hrule height.#2pt}}}
\def\square{\mathord{\dalemb{4.9}{5}\hbox{\hskip1pt}}}

\font\fourteentt=cmtt10 scaled \magstep2
\font\fourteenbf=cmbx12 scaled \magstep1
\font\fourteenrm=cmr12 scaled \magstep1
\font\fourteeni=cmmi12 scaled \magstep1
\font\fourteenssr=cmss12 scaled \magstep1
\font\fourteenmbi=cmmib10 scaled \magstep2
\font\fourteensy=cmsy10 scaled \magstep2
\font\fourteensl=cmsl12 scaled \magstep1
\font\fourteenex=cmex10 scaled \magstep2
\font\fourteenit=cmti12 scaled \magstep1
\font\twelvett=cmtt12 \font\twelvebf=cmbx12
\font\twelverm=cmr12  \font\twelvei=cmmi12
\font\twelvessr=cmss12 \font\twelvembi=cmmib10 scaled \magstep1
\font\twelvesy=cmsy10 scaled \magstep1
\font\twelvesl=cmsl12 \font\twelveex=cmex10 scaled \magstep1
\font\twelveit=cmti12
\font\tenssr=cmss10 \font\tenmbi=cmmib10
 
 \font\ninebf=cmbx9
\font\ninerm=cmr9  \font\ninei=cmmi9
\font\ninesy=cmsy9 \font\ninessr=cmss9
\font\ninembi=cmmib10 scaled 900
\font\eightit=cmti8 \font\eightsl=cmsl8
\font\eighttt=cmtt8 \font\eightbf=cmbx8
\font\eightrm=cmr8  \font\eighti=cmmi8
\font\eightsy=cmsy8 \font\eightex=cmex10 scaled 800
\font\eightssr=cmss8 \font\eightmbi=cmmib10 scaled 800
 
\font\sevenbf=cmbx7 \font\sevenrm=cmr7 \font\seveni=cmmi7
\font\sevensy=cmsy7 
\font\sevenssr=cmss9 scaled 778 \font\sevenmbi=cmmib10 scaled 700
 
 \font\sixbf=cmbx7 scaled 875
\font\sixrm=cmr6  \font\sixi=cmmi6
\font\sixsy=cmsy6 \font\sixssr=cmss8 scaled 750
\font\sixmbi=cmmib10 scaled 600
\font\fivessr=cmss8 scaled 625  \font\fivembi=cmmib10 scaled 500

\newskip\ttglue
\newfam\ssrfam
\newfam\mbifam

\mathchardef\alpha="710B
\mathchardef\beta="710C
\mathchardef\gamma="710D
\mathchardef\delta="710E
\mathchardef\epsilon="710F
\mathchardef\zeta="7110
\mathchardef\eta="7111
\mathchardef\theta="7112
\mathchardef\iota="7113
\mathchardef\kappa="7114
\mathchardef\lambda="7115
\mathchardef\mu="7116
\mathchardef\nu="7117
\mathchardef\xi="7118
\mathchardef\pi="7119
\mathchardef\rho="711A
\mathchardef\sigma="711B
\mathchardef\tau="711C
\mathchardef\upsilon="711D
\mathchardef\phi="711E
\mathchardef\chi="711F
\mathchardef\psi="7120
\mathchardef\omega="7121
\mathchardef\varepsilon="7122
\mathchardef\vartheta="7123
\mathchardef\varpi="7124
\mathchardef\varrho="7125
\mathchardef\varsigma="7126
\mathchardef\varphi="7127
\mathchardef\partial="7140

\def\fourteenpoint{\def\rm{\fam0\fourteenrm}
\textfont0=\fourteenrm \scriptfont0=\tenrm \scriptscriptfont0=\sevenrm
\textfont1=\fourteeni \scriptfont1=\teni \scriptscriptfont1=\seveni
\textfont2=\fourteensy \scriptfont2=\tensy \scriptscriptfont2=\sevensy
\textfont3=\fourteenex \scriptfont3=\fourteenex \scriptscriptfont3=\fourteenex
\def\it{\fam\itfam\fourteenit} \textfont\itfam=\fourteenit
\def\sl{\fam\slfam\fourteensl} \textfont\slfam=\fourteensl
\def\bf{\fam\bffam\fourteenbf} \textfont\bffam=\fourteenbf
\scriptfont\bffam=\tenbf \scriptscriptfont\bffam=\sevenbf
\def\tt{\fam\ttfam\fourteentt} \textfont\ttfam=\fourteentt
\def\ssr{\fam\ssrfam\fourteenssr} \textfont\ssrfam=\fourteenssr
\scriptfont\ssrfam=\tenmbi \scriptscriptfont\ssrfam=\sevenmbi
\def\mbi{\fam\mbifam\fourteenmbi} \textfont\mbifam=\fourteenmbi
\scriptfont\mbifam=\tenmbi \scriptscriptfont\mbifam=\sevenmbi
\tt \ttglue=.5em plus .25em minus .15em
\normalbaselineskip=16pt
\bigskipamount=16pt plus5pt minus5pt
\medskipamount=8pt plus3pt minus3pt
\smallskipamount=4pt plus1pt minus1pt
\abovedisplayskip=16pt plus 4pt minus 12pt
\belowdisplayskip=16pt plus 4pt minus 12pt
\abovedisplayshortskip=0pt plus 4pt
\belowdisplayshortskip=9pt plus 4pt minus 6pt
\parskip=5pt plus 1.5pt
\twelvefoot
\setbox\strutbox=\hbox{\vrule height12pt depth5pt width0pt}
\let\sc=\tenrm
\let\big=\fourteenbig \normalbaselines\rm}
\def\fourteenbig#1{{\hbox{$\left#1\vbox to12pt{}\right.\n@space$}}
\def\square{\mathord{\dalemb{6.8}{7}\hbox{\hskip1pt}}}}

\def\twelvepoint{\def\rm{\fam0\twelverm}
\textfont0=\twelverm \scriptfont0=\ninerm \scriptscriptfont0=\sevenrm
\textfont1=\twelvei \scriptfont1=\ninei \scriptscriptfont1=\seveni
\textfont2=\twelvesy \scriptfont2=\ninesy \scriptscriptfont2=\sevensy
\textfont3=\twelveex \scriptfont3=\twelveex \scriptscriptfont3=\twelveex
\def\it{\fam\itfam\twelveit} \textfont\itfam=\twelveit
\def\sl{\fam\slfam\twelvesl} \textfont\slfam=\twelvesl
\def\bf{\fam\bffam\twelvebf} \textfont\bffam=\twelvebf
\scriptfont\bffam=\ninebf \scriptscriptfont\bffam=\sevenbf
\def\tt{\fam\ttfam\twelvett} \textfont\ttfam=\twelvett
\def\ssr{\fam\ssrfam\twelvessr} \textfont\ssrfam=\twelvessr
\scriptfont\ssrfam=\ninessr \scriptscriptfont\ssrfam=\sevenssr
\def\mbi{\fam\mbifam\twelvembi} \textfont\mbifam=\twelvembi
\scriptfont\mbifam=\ninembi \scriptscriptfont\mbifam=\sevenmbi
\tt \ttglue=.5em plus .25em minus .15em
\normalbaselineskip=14pt
\bigskipamount=14pt plus4pt minus4pt
\medskipamount=7pt plus2pt minus2pt
\abovedisplayskip=14pt plus 3pt minus 10pt
\belowdisplayskip=14pt plus 3pt minus 10pt
\abovedisplayshortskip=0pt plus 3pt
\belowdisplayshortskip=8pt plus 3pt minus 5pt
\parskip=3pt plus 1.5pt
\tenfoot
\setbox\strutbox=\hbox{\vrule height10pt depth4pt width0pt}
\let\sc=\ninerm
\let\big=\twelvebig \normalbaselines\rm}
\def\twelvebig#1{{\hbox{$\left#1\vbox to10pt{}\right.\n@space$}}
\def\square{\mathord{\dalemb{5.9}{6}\hbox{\hskip1pt}}}}

\def\tenpoint{\def\rm{\fam0\tenrm}
\textfont0=\tenrm \scriptfont0=\sevenrm \scriptscriptfont0=\fiverm
\textfont1=\teni \scriptfont1=\seveni \scriptscriptfont1=\fivei
\textfont2=\tensy \scriptfont2=\sevensy \scriptscriptfont2=\fivesy
\textfont3=\tenex \scriptfont3=\tenex \scriptscriptfont3=\tenex
\def\it{\fam\itfam\tenit} \textfont\itfam=\tenit
\def\sl{\fam\slfam\tensl} \textfont\slfam=\tensl
\def\bf{\fam\bffam\tenbf} \textfont\bffam=\tenbf
\scriptfont\bffam=\sevenbf \scriptscriptfont\bffam=\fivebf
\def\tt{\fam\ttfam\tentt} \textfont\ttfam=\tentt
\def\ssr{\fam\ssrfam\tenssr} \textfont\ssrfam=\tenssr
\scriptfont\ssrfam=\sevenssr \scriptscriptfont\ssrfam=\fivessr
\def\mbi{\fam\mbifam\tenmbi} \textfont\mbifam=\tenmbi
\scriptfont\mbifam=\sevenmbi \scriptscriptfont\mbifam=\fivembi
\tt \ttglue=.5em plus .25em minus .15em
\normalbaselineskip=12pt
\bigskipamount=12pt plus4pt minus4pt
\medskipamount=6pt plus2pt minus2pt
\abovedisplayskip=12pt plus 3pt minus 9pt
\belowdisplayskip=12pt plus 3pt minus 9pt
\abovedisplayshortskip=0pt plus 3pt
\belowdisplayshortskip=7pt plus 3pt minus 4pt
\parskip=0.0pt plus 1.0pt
\eightfoot
\setbox\strutbox=\hbox{\vrule height8.5pt depth3.5pt width0pt}
\let\sc=\eightrm
\let\big=\tenbig \normalbaselines\rm}
\def\tenbig#1{{\hbox{$\left#1\vbox to8.5pt{}\right.\n@space$}}
\def\square{\mathord{\dalemb{4.9}{5}\hbox{\hskip1pt}}}}

\def\eightpoint{\def\rm{\fam0\eightrm}
\textfont0=\eightrm \scriptfont0=\sixrm \scriptscriptfont0=\fiverm
\textfont1=\eighti \scriptfont1=\sixi \scriptscriptfont1=\fivei
\textfont2=\eightsy \scriptfont2=\sixsy \scriptscriptfont2=\fivesy
\textfont3=\eightex \scriptfont3=\eightex \scriptscriptfont3=\eightex
\def\it{\fam\itfam\eightit} \textfont\itfam=\eightit
\def\sl{\fam\slfam\eightsl} \textfont\slfam=\eightsl
\def\bf{\fam\bffam\eightbf} \textfont\bffam=\eightbf
\scriptfont\bffam=\sixbf \scriptscriptfont\bffam=\fivebf
\def\tt{\fam\ttfam\eighttt} \textfont\ttfam=\eighttt
\def\ssr{\fam\ssrfam\eightssr} \textfont\ssrfam=\eightssr
\scriptfont\ssrfam=\sixssr \scriptscriptfont\ssrfam=\fivessr
\def\mbi{\fam\mbifam\eightmbi} \textfont\mbifam=\eightmbi
\scriptfont\mbifam=\sixmbi \scriptscriptfont\mbifam=\fivembi
\tt \ttglue=.5em plus .25em minus .15em
\normalbaselineskip=9pt
\bigskipamount=9pt plus3pt minus3pt
\medskipamount=5pt plus2pt minus2pt
\abovedisplayskip=9pt plus 3pt minus 9pt
\belowdisplayskip=9pt plus 3pt minus 9pt
\abovedisplayshortskip=0pt plus 3pt
\belowdisplayshortskip=5pt plus 3pt minus 4pt
\parskip=0.0pt plus 1.0pt
\setbox\strutbox=\hbox{\vrule height8.5pt depth3.5pt width0pt}
\let\sc=\sixrm
\let\big=\eightbig \normalbaselines\rm}
\def\eightbig#1{{\hbox{$\left#1\vbox to6.5pt{}\right.\n@space$}}
\def\square{\mathord{\dalemb{3.9}{4}\hbox{\hskip1pt}}}}

\def\vfootnote#1{\insert\footins\bgroup\footsuite
    \interlinepenalty=\interfootnotelinepenalty
    \splittopskip=\ht\strutbox
    \splitmaxdepth=\dp\strutbox \floatingpenalty=20000
    \leftskip=0pt \rightskip=0pt \spaceskip=0pt \xspaceskip=0pt
    \textindent{#1}\footstrut\futurelet\next\fo@t}
\def\hangfootnote#1{\edef\@sf{\spacefactor\the\spacefactor}#1\@sf
    \insert\footins\bgroup\footsuite
    \let\par=\endgraf
    \interlinepenalty=\interfootnotelinepenalty
    \splittopskip=\ht\strutbox
    \splitmaxdepth=\dp\strutbox \floatingpenalty=20000
    \leftskip=0pt \rightskip=0pt \spaceskip=0pt \xspaceskip=0pt
    \smallskip\item{#1}\bgroup\strut\aftergroup\@foot\let\next}
\def\footsuite{}
\def\twelvefoot{\def\footsuite{\twelvepoint}}
\def\tenfoot{\def\footsuite{\tenpoint}}
\def\eightfoot{\def\footsuite{\eightpoint}}
\catcode`@=12
\input epsf.tex
\def\arcsinh{\mathop{\rm arcsinh}\nolimits}

\def\crampest{\medmuskip = 1mu plus 1mu minus 1mu}
\def\uncramp{\medmuskip = 4mu plus 2mu minus 4mu}
\def\underbuildrel#1\under#2{\mathrel{\mathop{\kern0pt #2}\limits_{#1}}}
\def\im{{\rm i}}
\def\R{\rlap I\mkern3mu{\rm R}}
\def\C{\mkern1mu\raise2.2pt\hbox{$\scriptscriptstyle|$}\mkern-7mu{\rm C}}
\def\Z{Z}
\def\for{\lower6pt\hbox{$\Big|$}}
\def\subfor{\lower3pt\hbox{$|$}}
\def\gtlt{\mathrel{\raise4.5pt\hbox{\oalign{$\scriptstyle>$\crcr
$\scriptstyle<$}}}}
\twelvepoint
\nofirstpagenotwelve
\oneandahalfspace
\rightline{CTP TAMU--1/94}
\rightline{G\"oteborg ITP 94-3}
\rightline{Imperial/TP/93-94/13}
\rightline{hep-th/9401007}
\rightline{January 1994}

\vskip 2truecm
\centerline{\bf The Multivalued Free-Field Maps of Liouville and Toda
Gravities\hangfootnote{$^{\ast}$}{\tenfoot Research Supported in part by the
Commission of the European Communities under Contracts\nl SC1*0394--C,
SC1*--CT91--0674 and SC1*--CT92--0789.}}
\vskip 1.5truecm
\centerline{A. Anderson$^{1}$, B.E.W. Nilsson$^{2}$,
C.N. Pope$^{3}$\hangfootnote{$^{\dag}$}{\tenfoot Supported in part
by the U.S. Department of Energy, under
grant DE-FG05-91ER40633.} and K.S. Stelle$^{1}$}
\vskip 1truecm
\settabs\+\hskip 3cm &$^1$\ &\cr
\+&$^1$&{\it The Blackett Laboratory, Imperial College,}\cr
\+&&{\it London SW7 2BZ, U.K.}\cr
\+&$^2$&{\it Department of Physics, Chalmers University,}\cr
\+&&{\it Gothenburg, Sweden}\cr
\+&$^3$&{\it Center for Theoretical Physics, Texas A\&M University,}\cr
\+&&{\it College Station, TX 77843--4242, USA}\cr
\vskip 1.5truecm
\AB\singlespace
Liouville and Toda gravity theories with non-vanishing interaction
potentials have spectra obtained by dividing the free-field spectra for
these cases by the Weyl group of the corresponding $A_1$ or $A_2$ Lie
algebra. We study the canonical transformations between interacting and
free fields using the technique of intertwining operators, giving explicit
constructions for the wavefunctions and showing that they are invariant
under the corresponding Weyl groups. These explicit constructions also
permit a detailed analysis of the operator-state maps and of the nature of
the Seiberg bounds. \AE\oneandahalfspace

\np
\noindent{\bf 1. Introduction}
\bigskip

     Non-critical string theories achieve quantum consistency by
promoting the  local parameters of anomalous worldsheet symmetries to
dynamical fields  and then considering enlarged theories in which the
anomalies are canceled via the  dynamical effects of these additional
fields. The treatment of the Liouville mode of ordinary noncritical
string theory in such a fashion is by now familiar, dating from
Polyakov's original paper on the subject [1]; for recent reviews, see
Refs [2--5]. It should be kept in mind that the generic form of the trace
anomaly  relevant to worldsheet conformal transformations includes both a
term proportional to the Euler number of the worldsheet and also a
volume, or ``cosmological constant'' term [6]. Although the precise
coefficient of the latter depends on regularisation details of the
quantisation procedure, the important fact for our purposes is that this
coefficient is generically non-vanishing. The specific value of this
coefficient may in any case be altered by a constant shift in the
Liouville field, so no specific value is physically meaningful except for
its being nonvanishing. Therefore, one necessarily must deal with the
full nonlinear dynamics of the Liouville mode, and not just with its
contribution to the total central charge of the theory. Furthermore,
owing to the inessential nature of the Liouville potential's coefficient,
it is not appropriate to treat this nonlinearity in a perturbative
fashion.

     Extensions of bosonic string theories to models with worldsheet $W$
symmetries give rise to non-critical behaviour generalising that of the
Liouville case. The possible Lagrangians for the compensating modes
incorporated into the theory in order to cancel the anomalies may be
determined by consistency considerations. Similarly to the Liouville case,
the compensating modes taken together with the original matter fields must
form a realisation of the $W$-algebra with the specific value of the
central charge that is needed for overall cancellation of the anomalies.
The central-charge contributions from these modes, needed also for the
Virasoro subalgebra, may be obtained already at the ``classical'' level by
including background-charge terms $\int d^2\sigma\sqrt{\hat g}R(\hat g)\vec
Q\cdot\vec\varphi$ into the Lagrangian for the compensating modes
$\vec\varphi$, where $\hat g_{ij}$ is the fiducial worldsheet metric. For
gauge fields coupled to the higher-spin currents of the $W$ algebra, there
are analogous background-charge terms. The $W$-algebra transformations of
the  $\vec\varphi$ may be obtained in the operator formalism by taking
commutators with the $W$ generators. The interaction potential terms for
the $\vec\varphi$ must be left invariant by these transformations.
Demanding also that the classical limit as $\hbar\rightarrow 0$ be smooth,
one can determine the forms of the allowed potentials.

     To be specific, consider the Euclidean action for a Liouville field
arising as a result of the conformal anomalies:
$$
I={1\over 4\pi}\int d^2\sigma \sqrt{\hat g}\left(\ft12(\hat\nabla\varphi)^2
+  Q\varphi R(\hat g) + {4\over\gamma^2}e^{\gamma\phi}\right).\eqno(1.1)
$$
{}From this, one obtains the stress tensor by varying with respect to the
fiducial metric $\hat g_{ij}$, giving the standard result for the
chiral (holomorphic) component, where $z=e^{\im\sigma+\tau}$,
$$
T_{zz}=T=-\ft12(\partial\varphi)^2-Q\partial^2\varphi.\eqno(1.2)
$$
Using free-field commutation rules\rlap,\footnote{$^{\ast}$}{\tenfoot
Strictly speaking, a satisfactory derivation of the form of the potential
terms in non-critical string theories should be done on the basis of
consistency requirements within the interacting theory including the
potential. By requiring Virasoro covariance, this has been carried out in
standard non-critical string theory [7--12], giving the same result as the
free-field conformal-weight analysis summarised here. The analogous full
consistency analysis of the potentials for non-critical $W$-strings remains
an open problem.} one obtains the conformal weight of the potential
$e^{\gamma\varphi}$, $\Delta=-\ft12\gamma(\hbar\gamma+2Q)$. Thus the
condition for conformal invariance, that the potential term have weight
(1,1), is
$$
\gamma={1\over\hbar}\left(-Q\pm\sqrt{Q^2-2\hbar}\right),\eqno(1.3)
$$
where we have shown explicitly factors of $\hbar$ in order to discuss the
classical limit. Demanding that $\gamma$ have a smooth limit (to the
classical value $-Q^{-1}$) as $\hbar\rightarrow 0$, since anomaly
cancellation must be carried out order by  order in $\hbar$, one must
select the $+$ sign in (1.3) (for $Q>0$).

     In the case of non-critical $W_3$ strings, two compensating-mode
fields ($\varphi_1$,$\varphi_2$) are needed. The chiral stress tensor in the
presence of background charges ($Q_1$,$Q_2$) becomes
$$
T=T_1+T_2=\Big[-\ft12(\partial\varphi_1)^2-Q_1\partial^2\varphi_1\Big] +
\Big[-\ft12(\partial\varphi_2)^2-Q_2\partial^2\varphi_2\Big].\eqno(1.4)
$$
As in the Liouville case, potential terms $e^{\vec\gamma\cdot\vec\varphi}$
must have weight (1,1) under the (left, right) Virasoro transformations.
Using free-field commutation rules, the weight of an exponential
$e^{\vec\gamma\cdot\vec\varphi}$ is now given by
$\Delta=-\ft12\vec\gamma\cdot(\hbar\vec\gamma+2\vec Q)$, so setting
$\Delta=1$ yields an equation relating $\gamma_1$ and $\gamma_2$; this may
be solved to give
$$
\gamma_1={1\over\hbar}\left(-Q_1\pm\sqrt{Q_1^2-2\hbar-\hbar^2\gamma_2^2 -
2Q_2\gamma_2\hbar}\right).\eqno(1.5)
$$
Demanding that $\gamma_1$ have a smooth limit as $\hbar\rightarrow 0$
requires that we pick the $+$ sign in (1.5).

     In addition, one must require invariance of $\int d^2\sigma\sqrt{\hat g}
e^{\vec\gamma\cdot\vec\varphi}$ under the spin-3 transformations as well.
This may be achieved by requiring that the operator product of the
chiral spin-3 current [13]
$$
W=\ft13(\partial\varphi_1)^3+Q_1\partial\varphi_1\partial^2\varphi_1 +
\ft13Q_1^2\partial^3\varphi_1 + 2\partial\varphi_1T_2 + Q_1\partial
T_2\eqno(1.6)
$$
with $e^{\vec\gamma\cdot\vec\varphi}$ give a total
$\partial=\partial/\partial z$ derivative. Requiring this in the case of the
minimal $W_3$ string with only the two compensating modes $\varphi_1$ and
$\varphi_2$ present yields four solutions for $\gamma_2$, using the $+$
solution chosen in (1.5). In order for $T$ and $W$ to form a realisation of
the $W_3$ algebra, one must have $Q_1=\sqrt3Q_2$; the four solutions for
$\gamma_2$ are then
$$\eqalignno{
\gamma_2&={1\over\hbar}\left(-Q_2+\sqrt{-\ft12\hbar+
\ft52Q_2^2\mp\ft32Q_2\sqrt{Q_2^2-2\hbar}}\right)&(1.7a)\cr
\gamma_2&={1\over\hbar}\left(-Q_2\pm
\sqrt{Q_2^2-2\hbar}\right).&(1.7b)\cr}
$$
Requiring again a smooth classical limit as $\hbar\rightarrow 0$ eliminates
two of the above four possibilities, leaving just the $-$ solution of
(1.7$a$) and the $+$ solution of (1.7$b$). For a $W_3$ string with only two
scalar fields $\varphi_1$ and $\varphi_2$, the conditions for anomaly
cancellation require $Q_1=\sqrt{49\over8}$, implying
$Q_2=\sqrt{49\over24}$. Consequently, the allowable potentials for the
two-scalar $W_3$ string are [14]
$$
\eqalignno{
V_1&=e^{-\ft37Q_1\varphi_1+\ft37Q_2\varphi_2}&(1.8a)\cr
V_2&=e^{-\ft67Q_2\varphi_2}.&(1.8b)\cr}
$$

     In the context of free-field $W_3$ conformal field theory, the
potentials $V_1$ and $V_2$ may, upon replacing the field variables
$\varphi_1$ and  $\varphi_2$ by field operators, be viewed as screening
currents; indeed, our derivation of their structure used precisely this
relation. The other two solutions of (1.7), which we have rejected as
interaction potentials for the Lagrangian on the grounds of not having a
smooth classical limit, also give rise to screening currents in the
free-field conformal field theory. For reference, we give their
structure as well: $$\eqalignno{
\tilde V_1&=e^{-\ft47Q_1\varphi_1+\ft47Q_2\varphi_2}&(1.9a)\cr
\tilde V_2&=e^{-\ft87Q_2\varphi_2}.&(1.9b)\cr}
$$

     In this paper, we shall principally be concerned with the
two-scalar $W_3$ string without further matter fields. If one includes
extra matter fields, one must make a choice between the multi-scalar
construction of Refs\ [15,16], which proceeds by incorporating the extra
scalars into $T_2$ in (1.4, 1.6) and changing $Q_2$ so as to maintain the
effective central charge for $T_2$ equal to $\ft{51}2$, or the
construction of Ref.\ [17], which adds the matter fields in a separate
$W_3$ representation in such a fashion as to make an overall nilpotent
BRST operator for the whole theory. The free-field multi-scalar
construction of Refs [16,14] gives rise to two screening currents only, of
which one fails to have a smooth classical limit; the remaining one is
$V_2$. In the case of the construction of Ref.\ [17], we find that both
$V_1$ and $V_2$ in (1.8) generalise to give screening charges with smooth
classical limits.  In that case, one has background charges $Q_1$ and
$Q_2$ for the ``Toda'' sector, and background charges $Q_1^M$ and $Q_2^M$
for the ``matter'' sector.  The screening charges have momenta only in
the Toda sector, and are given by vertex operators
$$
\tilde V=e^{p_1\varphi_1+p_2\varphi_2}, \eqno(1.10)
$$
where $p_1/Q_1=-p_2/Q_2= -\ft12+\ft12 \sqrt{1-6/Q_1^2}$ for the
generalisation of $V_1$, and $p_1=0$, $p_2/Q_2=-1+\sqrt{1-6/Q_1^2}$ for the
generalisation of $V_2$.  As before there are two more screening charges,
corresponding to the other sign of the square root $\sqrt{1-6/Q_1^2}$; these
generalise the screening charges given by (1.9) that do not have smooth
classical limits.

     With the potentials (1.8) included into the Lagrangian, the two-scalar
$W_3$ string becomes an open Toda chain theory [18]. In the rest of this
paper, we shall study the spectra and dynamics of the Liouville and Toda
cases, emphasising the r\^ole of the center-of-mass zero modes that must be
treated non-perturbatively. We shall, as in many treatments of the
Liouville case [2--5], be concerned largely with the ``minisuperspace''
approximation, in which only the zero mode is kept, and no field-theory
oscillator excitations are considered. A new technique that we shall bring
to these much-studied questions is the approach of canonical
transformations for exactly-integrable systems {\it via} intertwining
operators, which has been elaborated in Refs [19--22]. The intertwining
operator is a quantum-mechanical implementation of a canonical
transformation, written in such a way as to give an explicitly-calculable
action on wavefunctions.

     In section two, we shall examine carefully the canonical transformation
in the Liouville case in order to set the stage for our further discussion.
For our purposes, an important fact will be that the canonical
transformation between the Liouville theory and its associated free-field
theory has a twofold branch structure. The branches of the transformation
in the general Toda case carry a representation of the Weyl group for the
underlying Lie algebra of the Toda theory, which in the Liouville case is
just $A_1$, with Weyl group $Z_2$. Next, we shall introduce the quantum
intertwining operator for the Liouville case, showing how the wavefunction
for the Liouville zero mode can be obtained, with the Weyl-multiplet
structure of the intertwining operator being reflected in the Weyl-group
symmetry of the wavefunction. This makes explicit the two-to-one nature of
the relation between the free theory and the Liouville theory, with two
states related by the Weyl group in the free theory being mapped onto the
same Liouville state. This formalism will then be compared to other quantum
canonical transformations in the literature, which mainly focus on the
relation between interacting-theory and free-theory field operators. We
shall rederive in particular the known results for the Heisenberg
representation field operator in the case of Liouville quantum mechanics
[23,24].

     In section three, we shall compare the present intertwining-operator
canonical transformation to the ``operator-state map'' path integral that
has frequently been employed to obtain interacting-theory wavefunctions
from time-asymptotic plane-wave forms of wavefunctions. We shall find,
through a careful analysis of the Liouville case, that this specific
construction introduces an asymmetry between positive and negative
plane-wave momenta. This is the origin of the ``Seiberg bound'' for the
Liouville theory, which is here seen to be a rather specific property of
the operator-state map rather than an intrinsic feature of Liouville
theory itself. The operator-state map for the important case of imaginary
plane-wave momenta will be seen to produce in general a superposition of
Liouville wavefunctions, and not just a single Liouville eigenfunction.

     In section four, we shall turn our attention to the $A_2$ Toda
theory, which as we have seen above is related to the $W_3$ string.
Paralleling our Liouville discussion, we first shall examine the
classical canonical transformations between the interacting theory and
the related free theory, emphasising the Weyl-multiplet structure of the
transformation. An intertwining-operator derivation  will then be given
of the wavefunction for the Toda center-of-mass zero modes, giving a
Weyl-symmetric result. This will establish the six-to-one nature of the
map between free-theory states and Toda states.

     In the conclusion, we shall discuss the implications of our results
for the spectra of non-critical strings and $W$-strings.  Much of the
existing literature on Liouville gravity [25,7--12] has been concerned
only with the relation between free and interacting-theory states in the
ghost-vacuum sector. For such states, the ghosts can be completely
factored out of the discussion. On the other hand, it is now well-known
[26,27] that, in the free theories associated to pure Liouville gravity
or Liouville gravity plus matter, there is a far richer structure of
physical states involving excitations of the ghost fields. We shall see
that the excited-state spectra of free Virasoro and $W_3$ strings have a
the Weyl-multiplet structure that combines states at different ghost
numbers. This suggests that the ghosts enter in an essential way into the
full field-theoretic multivalued maps between the interacting and free
theories in these cases.

     In the appendix, we shall present variations of the
intertwining-operator technique that produce forms of the Toda-theory
wavefunctions that have previously appeared [28--30] in the
integrable-model literature.

\bigskip
\noindent
{\bf 2. Canonical transformations for Liouville gravity}
\bigskip

     As discussed in the introduction, for the non-critical  bosonic string
the ghost-vacuum sector of the physical spectrum is described  by a
Liouville field theory coupled to the matter system.  At the quantum
level, the study of Liouville field theory is a complicated problem which
has not  been completely solved.  However, many of the subtleties that
arise are  associated specifically with the zero mode of the Liouville
field. This can be studied in isolation using the so-called minisuperspace
approximation [2--5,31].  In this approximation, Liouville field theory
reduces to Liouville quantum mechanics.

     An important feature of Liouville theory is its integrability; there
is a B\"acklund transformation that maps it into a completely free
theory.  This transformation, which can be implemented either at the level
of the field theory or in the minisuperspace limit, renders the theory
solvable, but it  does not imply that it is trivial. It is not trivial
because this transformation,  although canonical, is not unitary.  In
particular it is a two-to-one map. Also, there is no vacuum state in
Liouville theory since the  zero-momentum limit of the physical states is
not normalizable.  Nevertheless, one can in principle solve the Liouville
theory by mapping back from results for correlation functions in the
associated free theory, using the canonical transformation.  Since this
transformation is not unitary, one must be careful to preserve the
Hilbert-space inner product when mapping from the free to the interacting
theory, by applying suitable normalization factors.  Although these steps
have not been completely carried out for the full field theory, we shall
show that one can implement them completely at the quantum-mechanical
level.\footnote{$^{\dag}$}{\tenfoot There has been a considerable body of
work on quantum B\"acklund transformations for Liouville theory ({\it e.g.},
[11,32,33]).  Much of this has been concerned with the relations
between free-theory and interacting-theory field operators.  Here, we are
mostly concerned with the mapping of wavefunctions between the free and
interacting theories.  We shall, however, discuss the transformation of
field operators from our point of view later in this section.}

     We shall principally be concerned with the zero-mode contribution to
the  Liouville Hamiltonian, corresponding to the minisuperspace
approximation. We may split up the Liouville field and introduce a
convenient rescaling by
$$\eqalignno{
\varphi(\tau,\sigma)&={2\over\gamma}q(\tau) +
\varphi^{\rm osc}(\tau,\sigma),&(2.1a)\cr
\oint d\sigma\varphi^{\rm osc}&=0.&(2.1b)\cr}
$$
Then, ignoring for the moment the background-charge term, the action for
the minisuperspace mode is that for Liouville quantum mechanics,
$$
\pi\gamma I\ {\underbuildrel{\rm mss}\under\longrightarrow}\ I_{\rm
L}=\ft12\int d\tau\,\left(\dot q^2 + e^{2q}\right).\eqno(2.2)
$$
The background-charge term affects the relation between the momentum of
this Liouville system and the momentum customarily used to classify
states in conformal field theory, which is the parameter $\alpha$ in a
vertex operator $e^{\alpha\varphi(z,\bar z)}$, where
$z=e^{\im\sigma+\tau}$. Owing to the presence of the background charge in
(1.2), the conformal transformation of $\varphi$ from the $(z,\bar z)$
variables to the $(\tau,\sigma)$ variables induces an extra term in the
transformation of $\partial_z\varphi$, {\it i.e.}
$\partial_z\varphi\rightarrow(\partial_w\varphi-Q)/z$, where
$w=\im\sigma+\tau$. As a consequence, the total momentum inserted by a
vertex operator $e^{\alpha\varphi(z,\bar z)}$ is shifted with respect to
the Liouville momentum [2--5]:
$$
\im p_\varphi=\alpha+Q.\eqno(2.3)
$$
Accordingly, we shall have to take this shift into account in comparing
Liouville quantum mechanics to non-critical string theory.

\bigskip
\noindent{\it 2.1 Classical Liouville Mechanics}
\medskip

     We begin our discussion of the canonical transformations to a free
theory  at the classical level, letting the time parameter be denoted now
by $t$. The Hamiltonian for the minisuperspace mode following from (2.2) is
$$
H_{\rm L}=\ft12(p^2 + e^{2q}),\eqno(2.1.1)
$$
where $p(t)$ is the momentum canonically conjugate to $q(t)$.
Hamilton's equations give the equations of motion $\dot q=p$ and $\dot
p=-e^{2q}$, so we have $\ddot q + e^{2q}=0$.  Thus, from (2.1.1) we have a
first integral $\dot q^2 + e^{2q}=2E$, where the energy $E$ is a constant.
Performing the remaining integral, we obtain
$$
q=-\ln\Big({1\over \tilde p} \cosh(\tilde p t)\Big). \eqno(2.1.2)
$$
Here $\tilde p$ is an arbitrary constant, related to $E$ by $E=\ft12 \tilde
p^2$; the second constant of integration has been absorbed into the choice
of origin for $t$.  The solution may be written in the form
$$
\eqalignno{
e^{-q}&={1\over\tilde p} \cosh \tilde q &(2.1.3a)\cr
p&=-\tilde p \tanh\tilde q, &(2.1.3b)\cr}
$$
where $\tilde q=\tilde p t$.  In fact, these equations can be viewed as a
canonical transformation from the interacting Liouville system, with
phase-space coordinates $(q,p)$, to a free system with phase-space
coordinates $(\tilde q,\tilde p)$ and Hamiltonian
$$
\tilde H_{\rm L}=\ft12\tilde p^2.\eqno(2.1.4)
$$
Under this transformation, the solution $\tilde p=$constant, $\tilde
q=\tilde p t$ for the free system is mapped to the solution (2.1.2) for
the Liouville  system.

     The free Hamiltonian (2.1.4) has a $Z_2$ symmetry under the discrete
canonical transformation
$$
(\tilde q,\tilde p)\longrightarrow (-\tilde
q,-\tilde p).\eqno(2.1.5)
$$
This $Z_2$ can be understood as the Weyl
group of the $A_1=SL(2,\R)$ algebra underlying the Liouville
theory\rlap.\footnote{$^{\ddag}$}{\tenfoot This Weyl-group symmetry is
also the discrete residuum of a full $SL(2,\R)$ invariance of the
B\"acklund transformation for Liouville field theory [7].} Combined with
the canonical transformation (2.1.3), this symmetry gives another form
for the canonical transformation from the free to the interacting theory,
namely
$$
\eqalignno{
e^{-q}&=-{1\over\tilde p}\cosh\tilde q &(2.1.6a)\cr
p&=-\tilde p \tanh\tilde q. &(2.1.6b)\cr}
$$
Note that (2.1.3) and (2.1.6) have different classical domains of
applicability. Eq.\ (2.1.3) only makes sense classically for $\tilde p>0$,
and (2.1.6) only makes sense for $\tilde p<0$. The transformation (2.1.3)
maps a right-moving solution of the free theory, having positive momentum
$\tilde p$ and moving from $\tilde q=-\infty$ at $t=-\infty$ to $\tilde
q=+\infty$ at $t=+\infty$, into a solution of the Liouville theory that
starts from $q=-\infty$ at $t=-\infty$, bounces off the potential and goes
to $q=-\infty$ again at $t=+\infty$. For a left-moving solution of the free
theory, we must use the second canonical transformation (2.1.6).  This then
maps into the {\it identical} solution of the Liouville theory. Combining
the domains of (2.1.3) and (2.1.6), we have essentially only {\it one}
canonical transformation  from the free to the interacting theory. One may
summarise this by writing (2.1.3) and (2.1.6) together as
$$
\eqalignno{
e^{-q}&={1\over|\tilde p|}\cosh\tilde q &(2.1.7a)\cr
p&=-\tilde p \tanh\tilde q. &(2.1.7b)\cr}
$$

     Upon performing an inverse canonical transformation from the
interacting  Liouville theory to the free theory, one encounters a branch
structure: there  is a choice, for given $(q,p)$, of obtaining either
$(\tilde q,\tilde p)$ or  $(-\tilde q,-\tilde p)$, corresponding to the
choice of inverting (2.1.3) or (2.1.6) above.  This branch structure is the
basic expression of the two-to-one nature of the  map from the free to the
interacting Liouville theory.

     A further reflection of the two-to-one nature of the Liouville
free-field map is that, if we consider the asymptotic limit of the
transformation (2.1.7), we see that when $q$ is large and negative we have
$q\rightarrow \pm \tilde q+\ln(2|\tilde p|)$, so {\it both} asymptotic
regions of the free theory map into the {\it same} asymptotic region of the
Liouville theory.

     This asymptotic limit in $q$ corresponds to the behaviour of the
classical solutions as $t\rightarrow\pm\infty$. Specifically, for
$t\rightarrow-\infty$, and $\tilde p>0$, we find that $e^{-\tilde
q}=e^{-\tilde p t}$ dominates $e^{\tilde q}=e^{\tilde p t}$, so from (2.1.3)
we find $p\rightarrow\tilde p$ and $e^{-q}\rightarrow(2\tilde
p)^{-1}e^{-\tilde q}$. Thus, the time-asymptotic momentum of the
interacting theory gives the value of the free momentum $\tilde p$. For
$t\rightarrow-\infty$ with $\tilde p<0$, $e^{\tilde q}$ dominates
$e^{-\tilde q}$ and (2.1.6) implies $p\rightarrow-\tilde p$,
$e^{-q}\rightarrow-(2\tilde p)^{-1}e^{\tilde q}$. It is thus clear that the
cases $\tilde p>0$ and $\tilde p<0$ give rise to the same Liouville motion
$q(t)$, but with an opposite identification of the free momentum $\tilde p$
in terms of the asymptotic $p(t)$ as $t\rightarrow-\infty$.

     The limit as $t\rightarrow+\infty$ may be handled similarly, giving
$e^{-q}\rightarrow \pm(2\tilde p)^{-1}e^{\pm\tilde q}$;
$p\rightarrow\mp\tilde p$ for $\tilde p\gtlt0$. Note that the
$t\rightarrow\pm\infty$ limits of the Liouville momentum $p(t)$ differ
precisely by a Weyl-group reflection, so the conservation of $\tilde p$ in
the free theory implies the conservation of asymptotic values of $p(t)$ {\it
modulo} the Weyl group.

\bigskip
\noindent{\it 2.2 Quantum Liouville Mechanics}
\medskip

     The canonical transformation between the classical Liouville and free
theories discussed in the previous subsection can be generalised
to the quantum level.  We shall use the technique of intertwining operators
[19--22], which implements the canonical transformation on quantum
operators and also on wavefunctions. The canonical transformation will be
generated by an intertwining operator $C$, which maps any quantum operator
$A$ in the interacting theory into the corresponding operator $\tilde A$ in
the free theory according to the rule:
$$
C A C^{-1}=\tilde A .\eqno(2.2.1)
$$
In particular, we have the transformation $C H_{\rm L} C^{-1}=\tilde H_{\rm
L}$ from the interacting to the free Hamiltonian.  As a consequence, wave
functions $\psi$  and $\tilde \psi$ of the interacting and free theories
are related by
$$
\psi=C^{-1}\tilde\psi.\eqno(2.2.2)
$$
In order to interpret the latter, it is necessary to break up the canonical
transformation into a sequence of elementary canonical transformations, each
of which has a well-defined action on wavefunctions.

     Starting from the Liouville Hamiltonian (2.1.1), we perform the
following  sequence of elementary canonical transformations:
\medskip

\settabs\+\hskip 2cm &[{\rm L1}]\qquad\qquad&${\cal P}_{\ln q}$:
\qquad\qquad &$q\mapsto p^{-1} q p=q+ \im p^{-1}$, \qquad&$p\mapsto q p$\cr

\+&[{\rm L1}]&${\cal P}_{\ln q}$:&$q\mapsto \ln q$,&$p\mapsto q p$\cr

\+&[{\rm L2}]&$\cal I$:&$q\mapsto p$,&$p\mapsto -q $\cr

\+&[{\rm L3}]&$p^{-1}$:&$q\mapsto p^{-1} q p=q+ \im p^{-1}$,&$p\mapsto  p$\cr

\+&[{\rm L4}]&${\cal P}_{\sinh q}$:&$q\mapsto \sinh q$,&$p\mapsto
{1\over \cosh q} p$\cr

\medskip
\noindent
Acting on a wavefunction $\psi(q)$, the operator ${\cal P}_{f(q)}$ [L1, L4]
associated with the point transformation $q\mapsto f(q)$ makes the
replacement ${\cal P}_{f(q)}\psi(q)=\psi(f(q))$.  The operator $\cal I$
[L2], associated with the discrete canonical transformation $(q,p)\mapsto
(p,-q)$, is realised on functions of $q$ as a Fourier
transform\footnote{$^{\S}$}{\tenfoot The contours of integration in the
integral transforms $\cal I$ and ${\cal I}^{-1}$ must be chosen so that
boundary terms vanish when integrating by parts in verifying the
transformations of $q$ and $p$. In particular, this can allow limits
of integration other than $\pm\infty$.},
${\cal I}\psi(q) ={1\over\sqrt{2\pi}}\int dq\, e^{\im k q}\psi(q)$.
Its inverse ${\cal I}^{-1}$ is realised as an
inverse Fourier transform.  Finally, when acting on a function of $q$, $p$
is realised by differentiation, $p\, \psi(q)=-\im {\partial\over\partial
q}\psi(q)$, and its inverse $p^{-1}$ [L3] is realised by integration and
multiplication by $\im$.

     Wrapping the stages [L1--L4] together, the full relation between the
interacting and the free variables may be written
$$\eqalignno{
e^{-q}&={1\over\tilde p}\cosh\tilde q&(2.2.3a)\cr
p&=-\tanh(\tilde q)\tilde p.&(2.2.3b)\cr}
$$
Note that, although these take a form similar to the classical canonical
transformation (2.1.3), Eq.\ (2.2.3) is a full quantum-mechanical relation,
and the indicated ordering of the operators $(\tilde q, \tilde p)$ in it is
essential.

      Taking into account the commutation relation $[p,q]=-\im$, one can
easily perform the sequence of canonical transformations [L1--L4] on the
Liouville Hamiltonian:
$$
\eqalign{
2H_{\rm L}&=p^2 + e^{2q}\cr
[{\rm L1}]\qquad\qquad\qquad&\mapsto (q p)^2 +q^2=q^2p^2-\im q p +q^2\cr
[{\rm L2}]\qquad\qquad\qquad&\mapsto p^2 q^2 +\im p q +p^2\cr
[{\rm L3}]\qquad\qquad\qquad&\mapsto p q^2 p +\im q p  +p^2 =(1+q^2) p^2
-\im q p=\Big[(1+q^2)^{\ft12} p\Big]^2\cr
[{\rm L4}]\qquad\qquad\qquad&= \tilde p^2 =2 \tilde H_{\rm L}.\cr}
\eqno(2.2.4)
$$

     The inverse intertwining operator $C^{-1}$ is given by
$$
C^{-1}={\cal P}_{e^q}\,{\cal I}^{-1}\, p\,
{\cal P}_{\arcsinh q}.\eqno(2.2.5)
$$
Using it, we obtain the Liouville wavefunction starting from an
eigenfunction $e^{\im k\tilde q}$ of the free Hamiltonian.
Evaluating\footnote{$^{\P}$}{\tenfoot A related derivation may be found
in the appendix of Ref.\ [20].} the sequence of operations in
$C^{-1}e^{\im k\tilde q}$, we  have, leaving a
$k$-dependent normalization factor $N_k$ undetermined for the moment,
$$
\psi_k(q)=N_k\,{\cal P}_{e^q}\,{\cal I}^{-1}\,p\,{\cal
P}^{-1}_{\ft12(q-q^{-1})}q^{\im k},\eqno(2.2.6)
$$
since ${\cal P}_{\arcsinh q}={\cal P}^{-1}_{\ft12(q-q^{-1})}{\cal
P}^{-1}_{e^q}$. Next, we let $u=\ft12(y-y^{-1})$ define $y(u)$ implicitly.
In order to reproduce the transformation ${\cal P}_{\arcsinh q}$, it is
necessary to choose as solution the branch $y=u+\sqrt{u^2+1}$.
Then we have
$$\eqalign{
\psi_k(q)&={-\im N_k\over\sqrt{2\pi}}{\cal P}_{e^q}\int_{-\infty}^\infty
du\,e^{-\im qu}{\partial\over\partial u}(y(u))^{\im k}\cr
&={kN_k\over\sqrt{2\pi}}\int_0^\infty dy\,e^{-\ft\im2 e^q(y-y^{-1})}y^{\im
k-1}.\cr}\eqno(2.2.7)
$$
Strictly speaking, this integral needs to be regularized,
{\it e.g.}\ by rotating its integration contour by an angle
$\epsilon$ into the lower half plane and by taking $y-y^{-1}$ to mean
$y+e^{-\im\pi}y^{-1}$. Evaluating the integral, we obtain [34]
$$
\psi_k(q)={2kN_k\over\sqrt{2\pi}}e^{\pi k\over 2}K_{\im
k}(e^q),\eqno(2.2.8)
$$
where $K_{\im k}$ is a modified Bessel function.

     Note that the wavefunction (2.2.8) is not yet correctly normalized.
This is more than just an issue of a missing factor of
${1\over\sqrt{2\pi}}$  for the initial free-field plane wave. The
intertwining operator $C$ correctly generates a canonical transformation at
the quantum level, but this  transformation was never required to be
unitary. Instead, we are making a more general similarity transformation of
the form (2.2.1) on operators in the theory. One may calculate the effect
of the transformation $C$ on the Hilbert-space inner product, which is not
invariant under $C$ because $C$ is not unitary. In this way, we may learn
the ($k$-dependent) factor needed to convert (2.2.8) into a
fully-normalized wavefunction. The final result after normalization is:
$$
\psi_k(q)={1\over\pi}\sqrt{2k\sinh(\pi k)}\,K_{\im
k}(e^q).\eqno(2.2.9)
$$

     This wavefunction satisfies, for all complex
values\footnote{$^{\ast}$}{\tenfoot The modified Bessel function $K_{\im
k}(z)$ is, for fixed argument $z\ne 0$, an entire function of $k$; the same
is true for the normalized wavefunction (2.2.9). Thus,
although integral representations of Bessel functions such as (2.2.7) may
have limited domains of convergence, the result (2.2.8) can be analytically
continued throughout the complex $k$ plane.} of $k$, the physically-required
boundary condition of falling away to zero as $q\rightarrow\infty$,
{\it i.e.}\ under the Liouville potential. Of course, the time-independent
Schr\"odinger equation is a second-order differential equation and should
have a second solution which may, from general experience with quantum
mechanics, be expected {\it not} to satisfy the boundary conditions as
$q\rightarrow\infty$. This second, physically-rejected, solution to
Schr\"odinger's equation may also be obtained {\it via} an alternative
intertwining-operator transformation starting from the plane-wave
eigenfunctions of the free Hamiltonian $\tilde H_{\rm L}$. The existence of
another transformation that maps $H_{\rm L}$ into $\tilde H_{\rm L}$ can
easily be seen in the sequence of transformations (2.2.4), where the $\cal
I$ transformation [L2] acts upon an expression $q^2p^2-\im qp+q^2$ that is
already a polynomial of even order and hence is invariant under an extra
sign-flip transformation $(q,p)\mapsto(-q,-p)$, immediately preceding [L2].
Consequently, the transformation to the free Hamiltonian $\tilde H_{\rm L}$
may equally well be done by interchanging $\cal I$ with ${\cal I}^{-1}$.
Although this has no effect on the sequence of stages (2.2.4) to the free
Hamiltonian, the replacement of ${\cal I}^{-1}$ by $\cal I$ in the inverse
transformation $$
C'^{-1}\equiv{\cal P}_{e^q}\,{\cal I}\, p\, {\cal P}_{{\rm
arcsinh}\,q}\eqno(2.2.10)
$$
leads\footnote{$^{\dag}$}{\tenfoot In order to obtain a solution
independent from (2.2.7), the integrand in (2.2.10) should be taken to be
$\exp\big(\ft12\im e^q(y+e^{-\im\pi}y^{-1})\big) y^{\im k-1}$. The choice
of contour needed for convergence involves subtleties that we will not
address in this paper; but {\it cf.}\ Ref.\ [34].} to an un-normalizable
wavefunction
$$
\psi'_k(q)\propto K_{\im k}(e^{\im\pi}e^q).\eqno(2.2.11)
$$
This may be expressed in terms of the independent modified Bessel function
$I$ by use of the relation
$$
K_{\im k}(e^{\im\pi}e^q)=e^{\pi k}K_{\im k}(e^q) - \im \pi I^{\im
k}(e^q).\eqno(2.2.12)
$$
Note that this solution of Schr\"odinger's equation fails rather
spectacularly to obey the boundary condition of dying away under
the potential, because $I_{\im k}(e^q)$ diverges asymptotically in the limit
$q\rightarrow\infty$ like $\exp(e^q-q/2)$. In the sequel, we shall confine
our attention to the canonical transformation effected by (2.2.4) and the
corresponding wavefunction (2.2.7, 2.2.9).

     The quantum Liouville wavefunction is invariant under the $Z_2$ Weyl
group of $SU(2)$, as is manifest in the fully-normalized form (2.2.9), since
the modified Bessel function $K_{\im k}$ is symmetric under
$k\rightarrow-k$. Thus, the two-to-one character of the free-field to
Liouville map that we encountered at the classical level is equally present
at the quantum level: both free-field wavefunctions $e^{\im k\tilde q}$ and
$e^{-\im k\tilde q}$ map to the same Liouville wavefunction. One can
equivalently express this in terms of a $Z_2$ branch structure for the
inverse map from the interacting wavefunction to the free wavefunctions:
one is then faced with a choice of mapping $\psi_k(e^q)$ either back to
$e^{\im k\tilde q}$ or to $e^{-\im k\tilde q}$.

     In order to set the stage for our later Toda discussion in section 4,
we note also that the Weyl-group symmetry can already be seen in the
un-normalized form (2.2.7) of the wavefunction. Let us first rewrite the
integral in a manifestly convergent form by rotating the contour of
integration so as to run along the negative imaginary axis, making at the
same time the change of variables $y=e^{-\im\pi/2}w$. We thus obtain
$$
\psi_{k}(q)={N_{k}ke^{\pi k\over2}\over \sqrt{2\pi}}\int_0^\infty
dw\,e^{-{1\over2}e^q(w+w^{-1})}\, w^{\im k-1}.\eqno(2.2.13)
$$
At this stage, we may make a change in the variable of integration
$w\rightarrow w^{-1}$, yielding
$$
\psi_{k}(q)={N_{k}ke^{\pi k\over2}\over \sqrt{2\pi}}\int_0^\infty
dw\,e^{-{1\over2}e^q(w^{-1}+w)}\,w^{-\im k-1}.\eqno(2.2.14)
$$
This change of integration variable thus has the same effect on the
integral as a Weyl-group transformation $k\rightarrow-k$. Consequently,
application of a Weyl-group transformation to the un-normalised
wavefunction affects only the prefactor $ke^{\pi k\over2}$ of the integral
in (2.2.14); upon normalization, this is exchanged for the Weyl-symmetric
factor $(2\pi)^{-1}(2k\sinh(\pi k))^{1/2}$.

     Note that the form (2.2.13) of the integral representation for
$\psi_k(q)$ makes it manifest that the wavefunction falls off rapidly in
the classically-forbidden region $q>>0$ under the Liouville-potential
``mountain.''  In fact, from the asymptotic behaviour of the $K$ Bessel
function, one sees that it falls off as $\exp(-e^q-\ft12 q )$ at large
positive $q$.

     It is also worth noting that the manifestly-convergent integral
(2.2.13) for the wavefunction may itself be obtained directly from an
intertwining-operator realisation of a canonical transformation. This
transformation differs from the classically-motivated one [L1--L4] by the
replacement of the final [L4] point transformation by $[{\rm L4}']$: ${\cal
P}_{-\im\cosh q}$. This may also be viewed as an additional shift of the
free-field variable, $\tilde q\mapsto\tilde q-\ft12\im\pi$. This additional
transformation is not appropriate for the variables of the classical theory,
which must be real, but there is no reason to exclude it at the quantum
level. Combining the elementary transformations into an overall canonical
transformation, one has
$$\eqalignno{
e^{-q}&={-\im\over\tilde p}\sinh\tilde q&(2.2.15a)\cr
p&=-\coth(\tilde q)\tilde p,&(2.2.15b)\cr}
$$
which should be compared with the classically-implementable
transformation (2.2.3).

     Unlike (2.2.3), the form (2.2.15) for the transformation of the quantum
operators ($\tilde p$, $\tilde q$) is manifestly Weyl-group symmetric. This
does not necessarily mean that the intertwining-operator transformation of
the wavefunction is Weyl-group symmetric. It does imply, however, that a
Weyl-group transformation of the wavefunction (2.2.13) produces a variation
that is at worst an overall rescaling. For intertwining operators $C$ and
$C'$ differing by the application of a Weyl-group transformation, the
invariance of (2.2.15) implies that the operator $C^{-1}C'$ must commute
with $\tilde p$ and $\tilde q$. Thus it follows that [21]
$$
C^{-1}=\mu(C')^{-1},\qquad \mu\in\C\eqno(2.2.16)
$$
so the wavefunctions can differ at most by a scale factor $\mu$.

     In non-critical string theory, the Liouville momentum of physical
operators and states is given complex values. In the limit as
$q\rightarrow  -\infty$, where the Liouville potential dies away, the
wavefunction (2.2.9) becomes asymptotically
$$
\psi_k(q)\underbuildrel{q\to-\infty}\under\longrightarrow-\sqrt{2k\over
\sinh(\pi k)}
\sin\big[k(q-\ln 2)\big].
\eqno(2.2.17)
$$
Thus, if we set $k=k_{\scriptscriptstyle\rm R} + \im
k_{\scriptscriptstyle\rm I}$, where $k_{\scriptscriptstyle\rm R}$ and
$k_{\scriptscriptstyle\rm I}$ are the real and imaginary parts of $k$, this
means that the asymptotic behaviour of the wavefunctions as $q\rightarrow
-\infty$ is of the form
$$
\psi_k(q)\sim e^{-|k_{\rm I}|(q-\ln 2)} e^{\im s(k_{\rm I})
k_{\rm R}(q-\ln  2)},\eqno(2.2.18)
$$
where $s(k_{\rm I})=k_{\scriptscriptstyle\rm I}/|k_{\scriptscriptstyle\rm
I}|$.  In particular, we see that the wavefunctions  diverge exponentially
at large negative $q$, regardless of the sign of the imaginary part of $k$.
This means that when one maps from the free theory, the Liouville
wavefunction diverges at large negative $q$ regardless of whether the free
wavefunction diverges exponentially at large negative or at large positive
$\tilde q$.

     In order to show the relation between the intertwining-operator
approach to canonical transformations as used in this paper and
the more familiar transformations between the operators of the free and
interacting theories, we may use our results to rederive the
Heisenberg-representation time evolution of the quantum-mechanical
Liouville operator $e^{-q}$ [23,24]. In the full Liouville field theory,
similar relations between interacting-theory and free-theory quantum
operators assume a considerable importance because there is no {\it a
priori} well-defined mode expansion of the interacting Liouville field to
use when one wishes to include the oscillator modes in going beyond
the minisuperspace approximation. Consequently, one needs to map the mode
expansion of the free field into one for the interacting field {\it via} a
canonical transformation, demanding in the process covariance under the
Virasoro algebra and also locality [7--12]. It remains an important task
for the future to develop the present quantum-mechanical
intertwining-operator techniques so as to give a treatment of these classic
problems in the full Liouville quantum field theory.

     In the Heisenberg representation, the operator $e^{-q(t)}$ evolves
according to
$$
e^{-q(t)}=e^{\im H_{\rm L}(t-t_0)}e^{-q(t_0)}e^{-\im H_{\rm
L}(t-t_0)},\eqno(2.2.19)
$$
{\it i.e.}\ it is propagated by the interacting-theory Hamiltonian. We may
use the intertwining operator $C={\cal P}_{\sinh q}\,p^{-1}\,{\cal
I}\,{\cal P}_{\ln q}$ and its inverse $C^{-1}={\cal P}_{e^q}\,{\cal
I}^{-1}\, p \,{\cal P}_{\arcsinh q}$ to write (2.2.19) as
$$
e^{-q(t)}=C^{-1}e^{\ft\im2\tilde p^2(t-t_0)}Ce^{-q(t_0)}C^{-1}e^{-\ft\im2
\tilde p^2(t-t_0)}C.\eqno(2.2.20)
$$

     In order to evaluate (2.2.20), we begin in the center of this nested
expression and first evaluate $Ce^{-q(t_0)}C^{-1}$. In evaluating the
free-field map for operators, the normalization issue that one must face in
evaluating the map for wavefunctions is not relevant, since normalization
factors cancel out in calculating $CAC^{-1}$. Performing the nested
components of the transformation in the sequence [L1--L4], we obtain the
result (2.2.3) for the  canonical pair of operators at time $t_0$:
$$\eqalignno{
e^{-q(t_0)}&={1\over
\tilde p(t_0)}\cosh \tilde q(t_0).&(2.2.21a)\cr
p(t_0)&=-\big(\tanh\tilde q(t_0)\big)\, \tilde p(t_0).&(2.2.21b)}
$$

     Evolving (2.2.21) forward in time using the free-variable evolution
operator, we obtain
$$\eqalignno{
&e^{\ft\im2\tilde p^2(t-t_0)}{1\over \tilde p(t_0)}\cosh
\tilde q(t_0)\, e^{-\ft\im2
\tilde p^2(t-t_0)}={1\over\tilde p(t_0)}\cosh\Big(\tilde q(t_0)+\tilde
p(t_0)(t-t_0)\Big)&(2.2.22a)\cr
&=e^{-\ft\im2(t-t_0)}{1\over\tilde p(t_0)}\Big[\cosh
\tilde q(t_0)\cosh\Big(\tilde p(t_0)(t-t_0)\Big)+\sinh\tilde
q(t_0)\sinh\Big(
\tilde p(t_0)(t-t_0)\Big)\Big],\cr&&(2.2.22b)\cr}
$$
where we have used the Baker-Campbell-Hausdorff formula
$e^{A+B}=e^Ae^{-\ft12[A,B]}e^B$, which holds when $[A,[A,B]]=0=[B,[A,B]]$.
Completing the nested sequence of transformations of Eq.\ (2.2.20) by
applying $C^{-1}$ on the left and $C$ on the right, we end up with the
Heisenberg representation result of Refs [23,24].
$$
e^{-q(t)}=e^{-q(t_0)-\ft\im2(t-t_0)}\left(\cosh\left(\sqrt{2H_{\rm
L}}(t-t_0)\right)-p(t_0){1\over\sqrt{2H_{\rm L}}}\sinh\left(\sqrt{2H_{\rm
L}}(t-t_0)\right)\right),\eqno(2.2.23)
$$
where we have used the quantum relation $\sqrt{2H_{\rm
L}}=e^q\Big(1+(e^{-q}p)^2\Big)^{1/2}$, which may be checked by
verifying that its square correctly yields $2H_{\rm L}=p^2+e^{2q}$.

     To conclude this section, we compare the present intertwining-operator
approach to that of a related implementation of canonical transformations
[33], which  emphasises the r\^ole of the classical generating function
$F(q,\tilde q)$. At the classical level, the momenta can be derived from the
transformation formulas
$$
p={\partial F\over \partial q},\qquad \tilde p=-{\partial F\over
\partial\tilde q};\eqno(2.2.24)
$$
these equations implicitly determine $(\tilde q, \tilde p)$ in terms
of $(q,p)$. For the full Liouville field theory, the generating function is
$$
F(\phi, \tilde\phi)=\int
d\sigma\,(\phi\,\partial_\sigma\tilde\phi-e^\phi\sinh\tilde\phi),
\eqno(2.2.25)
$$
which in the minisuperspace limit $\phi(\tau,\sigma)\rightarrow q(t=\tau)$
becomes $F(q,\tilde q) = -e^q\sinh\tilde q$ and consequently (2.2.24)
reproduces our classical canonical transformation (2.1.3).

     As we have noted earlier, the full quantum canonical transformation
(2.2.3) takes a form similar to the classical transformation (2.1.3), but
it is essential to keep track of operator ordering at the quantum level.
Starting from (2.2.24), it is not always clear how one should generalise
the transformation generated by $F(q,\tilde q)$ to the quantum case. One
way to do this [33] is to require that both the interacting-theory and
free-theory Hamiltonians, considered as differential operators in $q$ and
$\tilde q$, have the same action on $e^{\im F(q,\tilde q)}$:
$$
H_{\rm L}(q,p)\,e^{\im F(q,\tilde q)}=\tilde H_{\rm L}(\tilde q,\tilde
p)\, e^{\im F(q,\tilde q)}.\eqno(2.2.26)
$$
That this holds for the Liouville generator (2.2.25) is something of an
accident; in general, one might expect $F(q,\tilde q)$ to acquire
quantum corrections, and to reduce to the classical expression only in the
limit $\hbar\rightarrow 0$.

     Once one has found a functional $F(q,\tilde q)$ with the property
(2.2.26), one can relate the interacting-theory and the free-theory
wavefunctions. If one starts from an eigenstate $\tilde\psi(\tilde q)$ of
the free Hamiltonian $\tilde H_{\rm L}$ with eigenvalue $E$ and constructs
the wavefunction
$$
\psi_{\rm L}(q)=\int d\tilde q\, e^{iF(q,\tilde q)}
\tilde\psi(\tilde q),\eqno(2.2.27)
$$
a consequence of (2.2.26) is that (2.2.27) is an eigenfunction of $H_{\rm
L}$ with the same eigenvalue $E$. At this point, the correspondence to our
intertwining-operator formulation becomes clear. The transformation
(2.2.27) should be compared to the application of $C^{-1}$ (2.2.5) to a
free wavefunction, yielding the relation
$$
C^{-1}=\int d\tilde q\, e^{iF(q,\tilde q)}.\eqno(2.2.28)
$$
This relation may be confirmed explicitly by a change of variables in
(2.2.7). Thus the procedure of Ref.\ [33] is an alternative way of looking
at the map from the free-theory wavefunction to the interacting-theory
wavefunction. The disadvantage of this procedure as compared to the
intertwining-operator approach adopted in this paper is that, by
concentrating on the accumulated form of the canonical transformation on
wavefunctions, one loses sight of the delicate operator orderings in the
quantum transformation of the canonical operator pair $(q,p)$, which
we obtained in (2.2.3). Also, if it had turned out that the classical
generator (2.2.25) had not itself satisfied (2.2.26), it would not have been
clear how to add $\hbar$-dependent corrections in order to make it do so.

\bigskip
\noindent {\bf 3. The operator-state map and the Seiberg bound}
\bigskip

     In the previous section, we explored the
quantum implementation of the canonical transformation between interacting
and free-field Liouville wavefunctions. At this point, we wish to explain
how this procedure differs from a construction frequently given for
interacting-theory wavefunctions in the string-theory literature [2--5].
This latter construction is motivated by the relation between operators and
states in conformal field theory. For a holomorphic conformal-field
operator ${\cal O}\big(\varphi(z)\big)$ in a free-field theory with a
state-space vacuum $|0\rangle$, a physical state associated to
${\cal O}\big(\varphi(z)\big)$ is obtained by taking the limit
$\lim_{z\rightarrow0}{\cal O}\big(\varphi(z)\big)|0\rangle$. In interacting
field theories such as Liouville
theory or Toda theory which do not have normalizable vacuum states in
their spectra, such a construction is not appropriate. Instead, many
authors use a path-integral construction of physical states
``associated'' to the operator $\cal O$. Without a vacuum, there is not
really a proper operator formalism for creating physical states. In the
path-integral formulation, $\cal O$ is not actually an operator, but
becomes instead a functional of the integration variables in the path
integral. Using now and for the rest of this section a Minkowskian
world-sheet signature, the interacting-theory wavefunction related to
${\cal O}$ is given by the construction
$$
\psi_{\cal O}\big(\varphi(\sigma)\big)=\underbuildrel{{\cal
D};\,\xi\subfor_{\partial{\cal
D}}=\varphi(\sigma)}\under{\int[d\xi(\sigma_i)]}e^{-\im I_{\rm L}}\,{\cal
O}(\xi).\eqno(3.1)
$$
The domain of integration $\cal D$ for $[d\xi]$ in (3.1) is over a surface
extending from negative temporal infinity, bounded by an end loop
$\partial{\cal D}$ on which Dirichlet boundary conditions
$\xi\subfor_{\partial{\cal D}}=\varphi(\sigma)$ are imposed. The factor of
$\cal O$ in (3.1) is a local insertion made at negative temporal infinity,
and the action $I_{\rm L}$ in the exponential weighting the path integral
is taken to be the action for the fully-interacting theory.

     In analogy with the vertex operators of free-field theory, the local
functional ${\cal O}$ in (3.1) is generally taken to have the form of a
free-field wave function. Specifically, for the Liouville theory, one takes
$e^{\im p\xi(\sigma^{\ast}_i)}$, where $\sigma^{\ast}_i$ is at negative
temporal infinity. In the implementation of the operator-state map for a
noncritical string theory, the action $I$ contains a background-charge term
as in (1.1). The integration over the constant mode of $\xi$ then produces
a shift in the momentum $p$ by a factor of $\im Q$, as we have already
discussed in (2.3). Aside from this shift and the related choice of a
relevant domain for the momentum $p$, the discussion of the operator-state
map in non-critical string theory may be reduced to a discussion in
Liouville field theory, and then in the minisuperspace approximation to
one in Liouville quantum mechanics.

     In the minisuperspace approximation, the field integration variable
$\xi(\sigma_i)$ in the path integral (3.1) reduces to a quantum-mechanical
variable $\xi(t)$, replacing as before $\tau$ by $t$ in the
quantum-mechanical case. In this limit, dropping the oscillator modes in
the decomposition (2.1), the path integral (3.1) becomes
$$
\psi_{\cal O}\big(q(t)\big)
=\lim_{t_0\rightarrow-\infty}\underbuildrel{t_0<t'<t;\ \xi\subfor_{t}=q(t)}
\under{\int[d\xi(t')]}\,e^{-\im
I_{\rm L}}\,{\cal O}\big(\xi(t_0)\big),\eqno(3.2)
$$
where the functional integration is over variables $\xi(t')$, $t_0< t'<t$,
with each $\xi(t')$ running from $-\infty$ to $\infty$, except for the
endpoint  $\xi(t)$ which is required to equal $q(t)$ by the Dirichlet
boundary condition. We recognise (3.2) as the standard Feynman
path-integral expression for the evolution up to time $t$ of a wavefunction
$\psi(q)$ starting out at time $t_0$ with the value ${\cal O}(q)$. In other
words, the operator-state map is nothing other than the application of the
evolution operator to the initial wavefunction ${\cal O}(q)$, {\it i.e.}
$$
\psi_{\cal O}\big(\varphi(t)\big)=\lim_{t_0\rightarrow-\infty}e^{-\im H_{\rm
L}(t-t_0)}{\cal O}\big(\varphi(t_0)\big).\eqno(3.3)
$$
Thus, the operator-state map derives a solution of the Liouville
Schr\"odinger equation from an initial time-asymptotic form of the
wavefunction, which may in turn be taken to be of a simple plane-wave form.

     The usefulness of this time-asymptotic construction is not {\it a
priori} obvious in an interacting theory such as the Liouville theory,
where the true quantum-mechanical stationary states are quite different
from plane waves.  In particular, as we saw in section 2, the true
stationary states die off very rapidly under the
potential ({\it i.e.}\ at large positive $q$), as $\exp(-e^q-\ft12 q)$, which
is
quite unlike the oscillatory behaviour of a plane wave state.  We shall see,
however, that at least in the case of an initial wavefunction $e^{\im
p\varphi(t_0)}$ with positive real momentum $p>0$, one does in fact generate
a single (un-normalized) Liouville wavefunction of the form (2.2.8). The
restriction
to real $p>0$ for the success of this construction constitutes the Seiberg
bound [4], which is thus seen to be a property of the operator-state map. We
shall see that the case with imaginary $p$, which is of importance to
noncritical string theory, actually produces a result that does not seem to
have been fully taken into account in the string-theory literature. It
should be emphasised that our present discussion of the operator-state map
does not in any way contradict the multivalued structure of the quantum
canonical transformations to free fields as discussed in section two. The
discussion here concerns the peculiarities of the operator-state map
construction, but does not really address the underlying structure of
Liouville theory itself. Our purpose in presenting this here in some detail
is to compare and
contrast the operator-state map to the more intrinsically-motivated
canonical transformation between interacting and free theories that we have
discussed earlier.

     Using the intertwining-operator realisation (2.2.5) of the canonical
transformation (2.2.1)\rlap, one can write (3.3) as
$$
\psi_{\cal O}\big(\varphi(t)\big)=\lim_{t_0\rightarrow-\infty}Ce^{-\im \tilde
H_{\rm L}(t-t_0)}C^{-1}{\cal O}\big(\varphi(t_0)\big).\eqno(3.4)
$$
Note that in both the free-Hamiltonian picture and in the
interacting-Hamiltonian picture, this construction has an ostensible
mismatch: from the point of view of (3.3), one is propagating a free-theory
wavefunction with the interacting-theory evolution operator. Alternatively,
upon noting that the application of $C^{-1}$ to
${\cal O}\big(\varphi(t_0)\big)$
gives an interacting-theory wavefunction as in (2.2.2), one sees that the
form (3.4) employs a free-theory evolution operator to propagate an
interacting-theory wavefunction. Consequently, the only hope for the
construction of (3.3, 3.4) to give a single Liouville-theory eigenfunction
is for just one of the infinity of exact Liouville-theory eigenmodes
arising in the decomposition of the initial wavefunction ${\cal
O}(\varphi)$ to persist in the infinite-time limit, and that all the other
modes have amplitudes that decay with time. We now shall see that this hope
is realised only in certain cases.

     There are two ways of explicitly evaluating the construction (3.4). One
can either decompose the initial wavefunction ${\cal O}(\varphi(t_0))$ into
normalized eigenstates (2.2.9) of the interacting Liouville Hamiltonian and
then evolve the latter forward in time by applying their corresponding phase
factors and by then recombining, or one can use our construction of the
Liouville intertwining operator (2.2.5) to explicitly calculate the Green
function for the theory. Both approaches lead to the same final integral
expression for the time-dependent wavefunction. Since the latter approach
is more in the spirit of our earlier discussion of the
Heisenberg-representation time evolution of the quantum operator
$e^{-q(t)}$ (2.2.19--2.2.23), this is the approach we shall present here.

     The Green function for the Liouville Schr\"odinger equation with a time
interval $\Delta t=t-t_0$ is given by
$$
G(z,w;\Delta t)=[C^{-1}e^{-\ft\im2\tilde p^2\Delta
t}C\delta(q-w)](z),\eqno(3.5)
$$
where the $q$ variable in $\delta(q-w)$ is transmuted into $z$ by the
operator. We begin by evaluating $e^{-\ft\im2\tilde p^2\Delta
t}C\delta(q-w)$, where
$C={\cal P}_{\sinh q}\,p^{-1}{\cal I}{\cal P}_{\ln q}$. In applying the
first two transformations within $C$ to the delta function, one needs to be
careful with the range of integration in the Fourier transform $I$.
Specifically, note that the transformation ${\cal I}{\cal P}_{\ln q}$
needs to be the inverse of ${\cal P}_{e^q}{\cal I}^{-1}$, which when acting
on functions $f(x)$ is realised by $[{\cal P}_{e^q}{\cal
I}^{-1}f](q)=(2\pi)^{-\ft12}\int_{-\infty}^\infty dx\,e^{-\im xe^q}f(x)$.
The inverse of this transformation, realised on functions $g(u)$ is given by
$(2\pi)^{-\ft12}\int_{-\infty}^\infty du\,e^ue^{\im xe^u}g(u)$. Writing this
in the form $[{\cal I}{\cal P}_{\ln q}g](x)$ by making the substitution
$q=e^u$, one obtains
$$
[{\cal I}{\cal P}_{\ln q}g](x)={1\over\sqrt{2\pi}}\int_0^\infty dq\,e^{\im
qx}g(\ln q).\eqno(3.6)
$$
The important point to note in (3.6) is that the lower limit of the $q$
integration is $0$ and not $-\infty$. Applying (3.6) to the integral
representation of the delta function
$\delta(q-w)=(2\pi)^{-1}\int_{-\infty}^\infty dk_1\,e^{\im k_1(q-w)}$, one
obtains
$$
[{\cal I}{\cal P}_{\ln q}\,\delta(q-w)](x)={1\over\sqrt{2\pi}}\int_0^\infty
dq\,e^{\im qx}\int_{-\infty}^\infty{dk_1\over2\pi}\,e^{-\im
k_1w}q^{\im k_1}.\eqno(3.7)
$$
Applying $e^{-\ft\im2\tilde p^2\Delta t}{\cal P}_{\sinh q}p^{-1}$ to
(3.7), one then obtains, using also $p^{-1}{\cal I}={\cal
I}q^{-1},$\crampest
$$
\eqalignno{
&[e^{-\ft\im2\tilde p^2\Delta t}C\delta(q-w)](z)\cr
&=\int_{-\infty}^\infty{dk_2\over\sqrt{2\pi}}\,
e^{\im zk_2}e^{-\ft\im2 k_2^2\Delta t}
\int_{-\infty}^\infty{dx\over\sqrt{2\pi}}\,
e^{-\im xk_2}\int_{-\infty}^\infty
{dk_1\over2\pi}\int_0^\infty{dq\over\sqrt{2\pi}}\,e^{-\im k_1w}q^{\im
k_1-1}e^{\im q\sinh x}\qquad&(3.8a)\cr
&=\int_{-\infty}^\infty
dk_1\int_{-\infty}^\infty{dk_2\over(2\pi)^\ft52}\,2^{\im k_1-1}e^{\ft12\pi
k_2}\Gamma\left(\im(k_1+k_2)\over2\right)\Gamma\left(\im(k_1-
k_2)\over2\right)e^{-\im k_1w}e^{\im k_2z-\ft\im2 k_2^2\Delta t}.&(3.8b)}
$$
\uncramp
The final stage of applying $C^{-1}$ may then be done using
$$
C^{-1}e^{\im k_2z}={2k_2\over\sqrt{2\pi}}e^{\ft\pi2 k_2}K_{\im
k_2}(e^z),\eqno(3.9)
$$
as derived earlier in (2.2.8). The final result for the Green function
is
$$
\eqalign{
&G(z,w;\Delta t)=\cr
&\ \int_{-\infty}^\infty
dk_1\int_{-\infty}^\infty{dk_2\over(2\pi)^3}\,2^{\im k_1}k_2e^{\pi
k_2}K_{\im k_2}(e^z)\Gamma\left(\im(k_1+k_2)\over2\right)\Gamma\left(\im(
k_1-k_2)\over2\right)e^{-\im k_1w}e^{-\ft\im2 k_2^2\Delta t}.\cr}\eqno(3.10)
$$

     Applying $\int_{-\infty}^\infty dw\, G(q,w;\Delta t)$ to an initial
plane-wave wavefunction at time $t_0$
$$
\psi\subfor_{{\cal O}_p}(w,t_0)=e^{\im pw},\eqno(3.11)
$$
one obtains the time-evolved wavefunction
$$
\psi\subfor_{{\cal O}_p}(q,t)= 2^{\im
p}\int_{-\infty}^\infty{dk\over(2\pi)^2}\, ke^{\pi k}K_{\im k}(e^q)
\Gamma\left(\im(p+k)\over2\right)\Gamma\left(\im(p-k)\over2\right)
e^{-\ft\im2 k^2\Delta t}.\eqno(3.12)
$$

     The behaviour of $\psi\subfor_{{\cal O}_p}(q,t)$ as $\Delta
t\rightarrow\infty$ may be evaluated by contour-integral methods. In doing
this, the point to note is that for $k=e^{-\im\pi/4}k_m$, the factor
$e^{-\im k^2\Delta t/2}$ becomes a representation of a delta function,
$e^{-k_m^2\Delta t/2}\rightarrow\sqrt{(2\pi/\Delta t)}\,\delta(k_m)$.
This causes the contributions to an integral of the form (3.12), but rotated
to the diagonal contour ${\rm Im}\,k=-{\rm Re}\,k$, to become concentrated
near the origin where the integrand vanishes, owing to the factor of
$k_m$. The rotated integral may be evaluated by steepest descents using
$$
\eqalign{
\int_{-\infty}^\infty dk_m\,k_mf(k_m)e^{-\ft12 k_m^2\Delta
t}&={1\over\Delta t}\int_{-\infty} ^\infty
dk_m\,f'(k_m)e^{-\ft12k_m^2\Delta t}\cr
&\rightarrow\sqrt{2\pi}(\Delta t)^{-\ft32}f'(0),\cr}\eqno(3.13)
$$
with $f(k_m)=\exp(\pi
e^{-\im\pi/4}k_m)K_{e^{\im\pi/4}k_m}(e^q)
\Gamma\big(\im(p+e^{-\im\pi/4}k_m)/2\big)
\Gamma\big(\im(p-e^{-\im\pi/4}k_m)/2\big)$. Noting that this expression
is $\exp(\pi e^{-\im\pi/4}k_m)$ times a factor even in $k_m$, we obtain the
large-time limit for the rotated integral
$$
{\rm I}\for_{\rlap{$\scriptstyle k=e^{-\im\pi/4}k_m$}}\
(q,\Delta t)\quad\underbuildrel{\Delta
t\rightarrow\infty}\under\longrightarrow{2^{\im
p-1}e^{-\ft14\im\pi}\over(2\pi)^2} \left[\Gamma(\ft\im2
p)\right]^2\left(2\pi\over\Delta t\right)^{\ft32}K_0(e^q).\eqno(3.14)
$$
Thus, the rotated integral falls off as $(\Delta t)^{-3/2}$ for large
$\Delta t$. Consequently, the value of $\psi\subfor_{{\cal O}_p}(q,t)$ in
the limit of large $\Delta t$ is given entirely by contributions from poles
of the integrand that become enclosed when joining the original integral
(3.12) to the rotated integral. These integrals are joined by two octant
arcs at infinity, $(3\pi/4)<\theta<\pi$ and $(7\pi/4)<\theta<2\pi$, on
which the $e^{-k^2\Delta t/2}$ factor causes the arc integrals to
vanish.

     The determination of the specific poles of the integrand that become
enclosed by the contour depends on the value of the momentum $p$ of the
initial wavefunction (3.11). The poles in (3.12) arise only from the
$\Gamma$ functions, since the rest of the integrand is an entire function
of $k$. The Gamma function $\Gamma\left(\im(p+k)\over2\right)$ has poles at
$k=-p+2\im n$, while the Gamma function $\Gamma\left(\im(p-k)\over2\right)$
has poles at $k=p-2\im n$, where $n\in\Z$, $n\ge0$.

     The easiest case to discuss (which we shall denote (I)) is that of
imaginary momentum $p=-\im \beta$, with  $\beta >0$, for which the initial
wavefunction takes the form
$$
\psi\subfor_{{\cal O}_{-\im \beta}}(q,t_0)=e^{\beta q}.\eqno(3.15)
$$
As can be seen from Figure 1, no poles are enclosed by the contour in this
case. Consequently, the integral (3.12) is given entirely by (3.14), and
falls off like $(\Delta t)^{-3/2}$ in this case.

    Thus when the momentum $p$ of the initial wavefunction is given by
$p=-\im \beta$ with $\beta >0$, there is no persistent Liouville
eigenfunction at late times.  This property has been expressed in the
literature by the statement that ``vertex operators $e^{\alpha \phi}$ with
$\alpha < -Q$ do not exist'' [5].\footnote{$^{\ddag}$}{\tenfoot Restated in
the conventions of this paper. Recall that the momentum of the vertex
operator $e^{\alpha\phi}$ is shifted by the background charge $Q$ relative
to the momentum of the initial wavefunction, and so
$\alpha=\ft\gamma2\beta-Q$, where $\gamma$ is a negative multiple of $Q$ as
given in (1.3); note also that $\varphi$ and $q$ are related as in (2.1a).}
The bound $\beta<0$ on the imaginary momenta of initial wavefunctions that
can leave persistent Liouville eigenfunctions at late times is known as the
``Seiberg bound'' [4].  Case (I), where there is no persistent eigenfunction,
is known as the ``anti-Seiberg'' case.
\bigskip
\centerline{\epsfbox{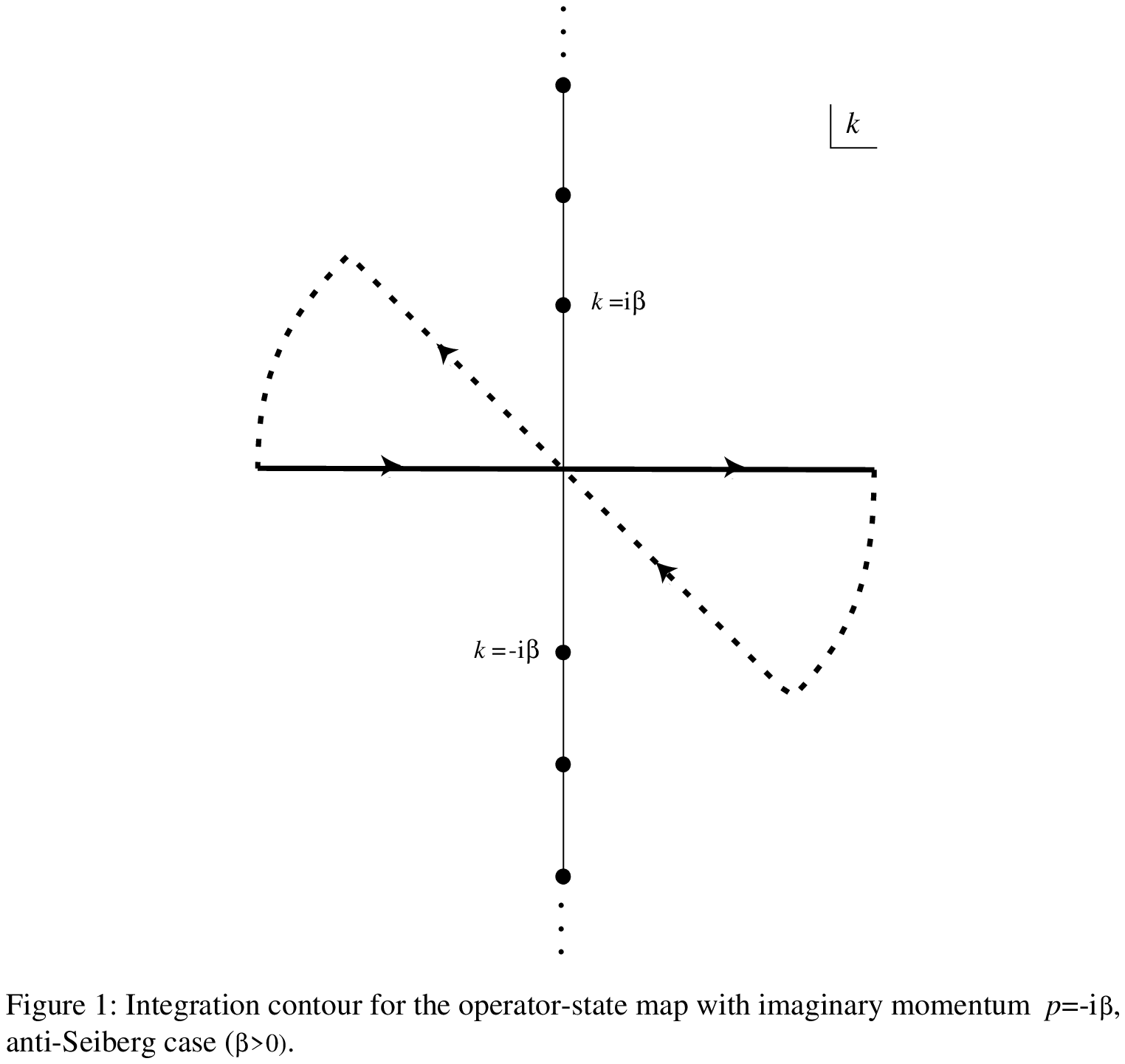}}

     The other cases of interest may be treated by analytic continuation in
$p$ from case (I), for which the integral in (3.12) is the most clearly
convergent. In making such analytic continuations, it is important not to
let a pole actually touch the integration contour. Thus, in order to treat
the cases of real momentum $p$, it is necessary to rotate $p$ from the
negative imaginary axis to an infinitesimal amount $\epsilon$ short of the
real axis, rotating either counterclockwise to obtain case (II): $p$
real, $p>0$, or clockwise to obtain case (III): $p$ real, $p<0$. These
cases are shown in Figure 2.

     As can be seen, in case (II), $p>0$, the $n=0$ Gamma function pole at
$k=-p$ is enclosed, as well as a number\footnote{$^{\S}$}{\tenfoot The
number of poles enclosed is $2[\ft12p]+2$, where $[x]$ is the integer part
of the real number $x$.} of $n>0$ resonance poles, depending on the value of
$p$. Since these poles have a time dependence with amplitude $e^{-2np\Delta
t}$, only the $n=0$ pole survives in the large $\Delta t$ limit. Thus, for
real $p>0$, the time-evolved wavefunction (3.12) does in fact settle down
to become a single (un-normalized) Liouville eigenfunction $K_{\im
p}(e^q)$ (once again, a single wavefunction occurs because $K_{\im p}$ is
symmetric in $p$).
\bigskip
\centerline{\epsfbox{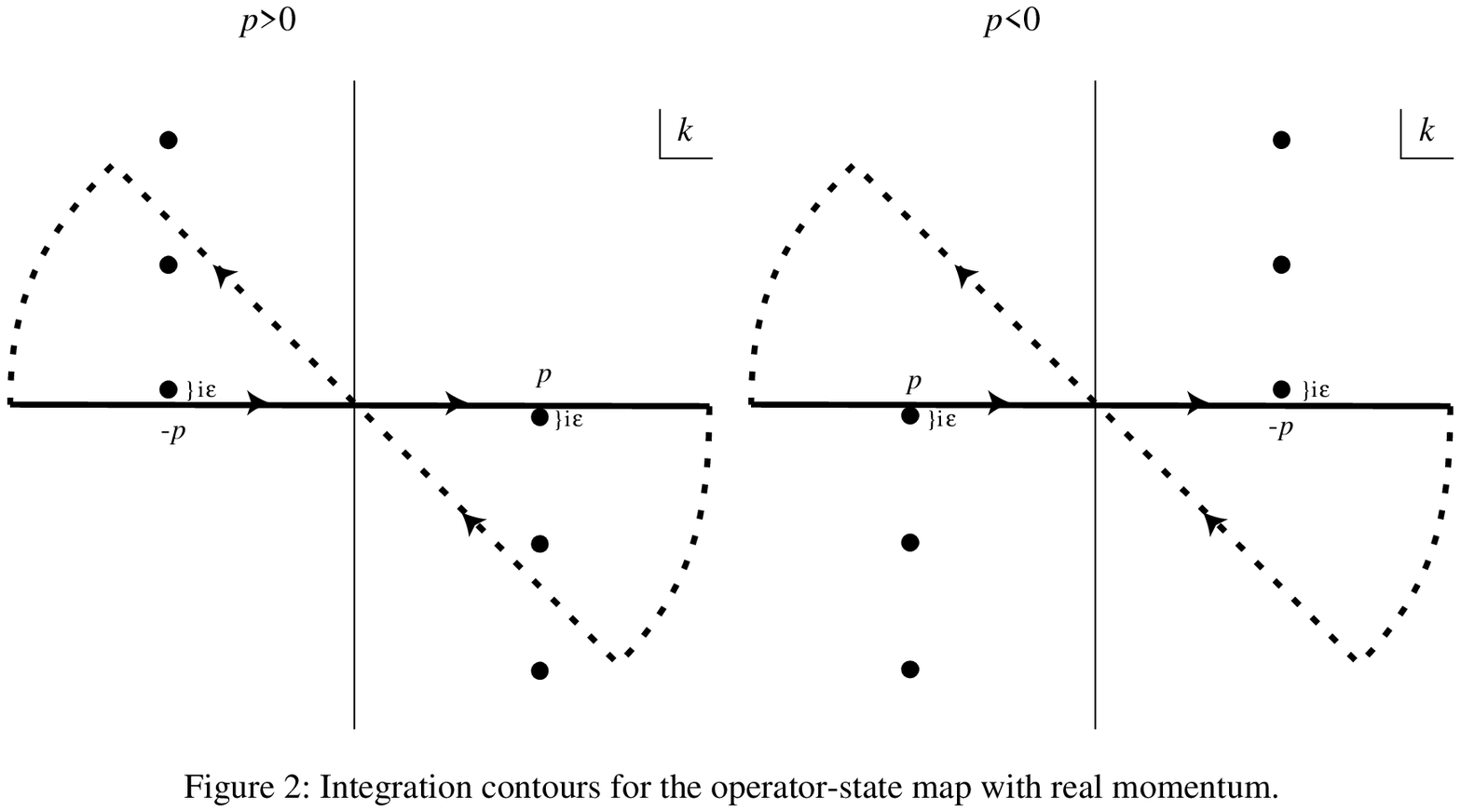}}

     The clockwise rotation from case (I) to case (III) leaves all
Gamma-function poles lying outside the integration contour. Note that the
requirement of stopping the rotation just $\epsilon$ short of bringing $p$
onto the real axis, in order to preserve convergence means that even the
$n=0$ pole is left outside the contour. Consequently, in case (III), of real
$p<0$, the only contribution to (3.12) in the large $\Delta t$ limit is
just (3.14), decaying like $(\Delta t)^{-3/2}$.

    The results for cases (II) and (III) taken together show that, starting
from an initial wavefunction with real momentum $p$, one only obtains a
persisting Liouville eigenfunction at late times if $p$ is positive. The
time-asymptotic nature of the operator-state map may be used to give a
heuristic picture of the difference between the real-momentum cases (II)
and (III). In case (II), if one starts from an initial wavefunction with
momentum directed toward the Liouville potential ``mountain'', the incoming
right-moving wave is reflected as time goes on, producing a left-moving
wave. At the same time, the wavefunction under the potential decays,
contributing to the left-moving wave. In the large $\Delta t$ limit,
the right-moving and left-moving waves combine to form a standing-wave
eigenfunction $\psi_p(q)$. In case (III), on the other hand, one starts
with a left-moving wave. As time goes on, the wavefunction under the
potential decays, contributing more to the left-moving wave. In this case,
however, there is no reflection and hence no contribution to any
right-moving wave. Thus, there is no way to produce a single standing-wave
eigenfunction. As a result, the wavefunction at all finite values of $q$
decays like
$(\Delta t)^{-3/2}$.
\bigskip
\centerline{\epsfbox{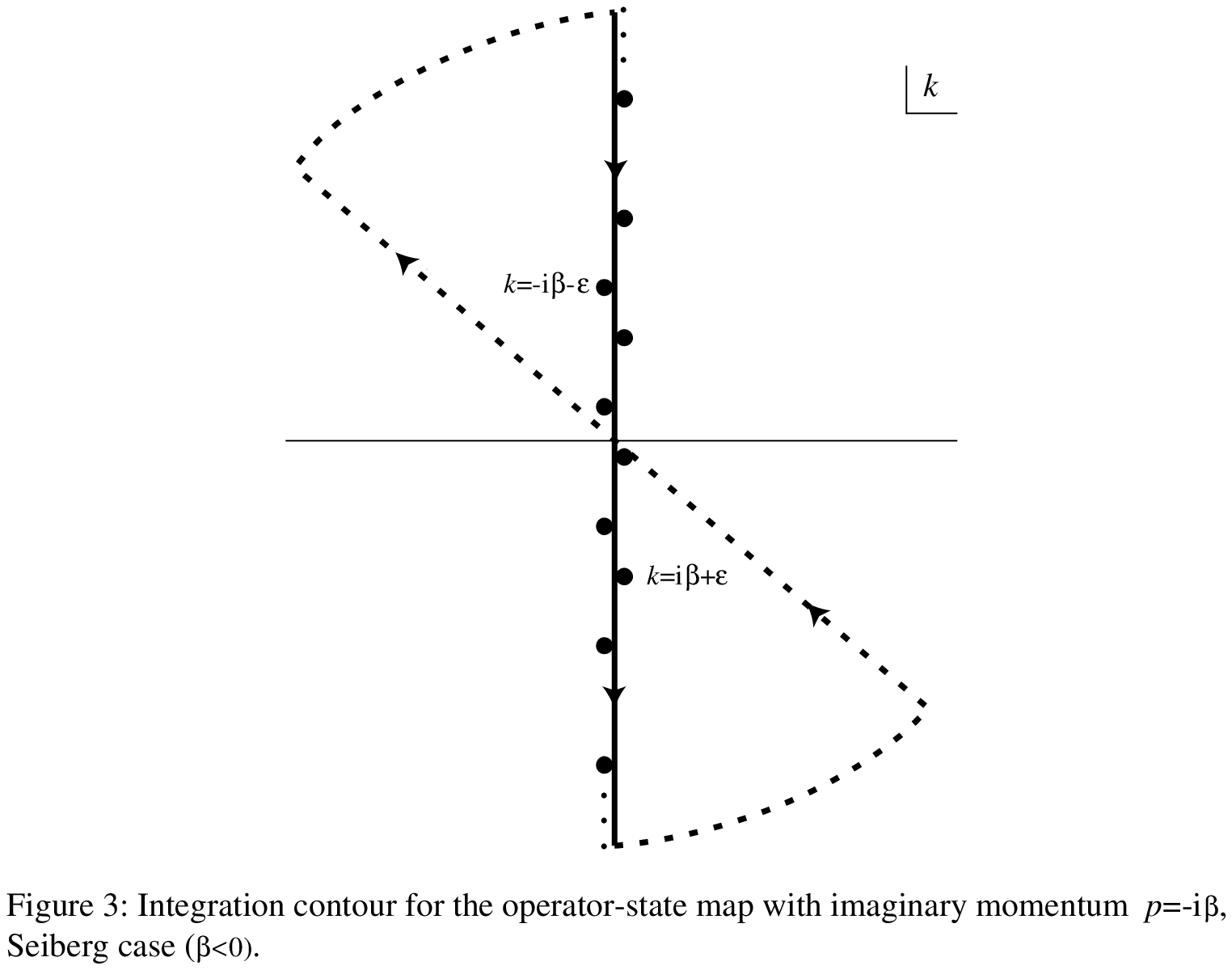}}

     The final case (IV), of imaginary $p=-\im \beta$, $\beta < 0$, has
been called the ``Seiberg'' case. It is not possible to reach this
configuration of the poles directly by starting from case (I) without having
some poles cross the real-axis contour of the original integral (3.12).
However, one may define this case by a two-step procedure of analytic
continuation. First, analytically continue $p=-\im \beta$ in case (I) so as
to give $p$ an infinitesimal negative real part $\epsilon$. Next, rotate the
contour of the integral (3.12) so as to make it run along the imaginary
axis, by adding quadrant arcs $\pi/2<\theta<\pi$ and
$(3\pi)/2<\theta<2\pi$, along which the integral vanishes. The shifted
poles and the rotated contour are shown in Figure 3. At this stage, one can
now analytically continue in $p$ parallel to the imaginary $k$ axis without
having poles cross the new contour. Evaluation of this redefined form of
the original integral (3.12) then proceeds as in cases (I), (II) and (III),
adding octant arcs with vanishing contributions and also the diagonal
integral along the contour
${\rm Im}\,k=-{\rm Re}\,k$, which again gives a decaying contribution
(3.14). As can be seen, a number of the Gamma-function poles are now
enclosed by this contour. Moreover, unlike in case (II), all of these poles
now have a pure-phase time evolution, since $(-\im k^2\Delta t)$ is purely
imaginary (up to infinitesimal terms from the regularization) for poles
lying along the imaginary axis. Thus, in addition to the expected Liouville
eigenfunction $K_{-\beta}(e^q)$, one obtains also a whole chorus of resonant
eigenfunctions\footnote{$^{\P}$}{\tenfoot The number of distinct persisting
Liouville eigenfunctions is $[-\ft12\beta]+1$.  Recall that $\beta$ is
negative here.} $K_{-\beta-2n}(e^q)$ with imaginary momenta $\im k_{\rm m}$
extending through positive values from $k_{\rm m}=-\beta$ down to zero, at
intervals $\Delta k_{\rm m}=2$.

     In the imaginary-momentum cases $p=-\im \beta$ (I, IV)
that are relevant in non-critical string theory, we have confirmed the
absence of persistent eigenfunctions from the operator-state map
construction as expected in the anti-Seiberg case $\beta>0$, but we get
more than we bargained for in the Seiberg case $\beta<0$: we find a
momentum-dependent number of modes persisting in the limit $\Delta
t\rightarrow\infty$.

     It is useful at this stage to summarize how the four cases (I--IV)
relate to one another. The integral (3.12) is most clearly defined in the
anti-Seiberg case (I). From this case, the others have been obtained by
analytic continuation in the momentum $p$, and also, for case (IV), in the
defining contour of the integral. One may visualize the lines of
poles of the Gamma functions in (3.12) as a ``stalagmite'' rising from the
floor with its tip located at $k=p$, and a ``stalactite'' descending from
the ceiling with its tip located at $k=-p$. In cases (I), (II) and (III),
these spikes do not pierce the real $k$ axis, which forms the boundary of
the region of convergence of (3.12) as originally defined. In these three
cases, one obtains a nonvanishing result at large times whenever there are
poles enclosed by the octant wedges of the closed contour of integration, as
shown in Figs 1 and 2. In case (IV), where $p=-\im\beta$ with $\beta<0$,
the stalagmite and stalactite have pierced the real $k$ axis, as shown in
Fig.\ 3. In consequence, one must be careful here in performing an analytic
continuation from any of the cases (I) (II) or (III). In particular, a
consistent analytic continuation to this case can be achieved by
deforming the defining contour of the integral (3.12) into one running
along the imaginary axis; this procedure avoids ever having a pole actually
touch the defining contour of integration. Note futher that the only cases
in which (3.12) can asymptotically settle down to have a time-independent
amplitude arise when some of the poles lie infinitesimally close to either
the real or imaginary $k$ axes.

     We conclude this section by comparing our findings on the
operator-state map with previous discussions in the literature.  When the
momentum is imaginary,  $k=\im k_{\scriptscriptstyle\rm I}$,  the
asymptotic form (2.2.18) of the Liouville eigenfunctions at large negative
$q$ ({\it i.e.}\ away from the potential) diverges exponentially,
regardless of the sign of $k_{\scriptscriptstyle\rm I}$.  Thus
(paraphrasing in the language of this paper) it was argued [31] that if an
initial wavefunction $e^{\beta q}$ had $\beta>0$, it could not give rise
to a persisting Liouville eigenfunction at late times.  Our results for
case (I) agree with this assertion.  As far as we are aware, however, it has
generally been assumed that in case (IV), where $\beta<0$, an initial
wavefunction $e^{\beta q}$ would result in a {\it single} persisting
Liouville eigenfunction $K_{-\beta}(e^q)$.  Our results show that this is
not generally what happens, and instead one gets a number of Liouville
eigenfunctions $K_{-\beta-2n}(e^q)$, with $n=0$, 1, 2, $\ldots
[-\ft12\beta]$.  In view of this finding, it is no longer clear that the
use of exponential vertex operators, which is very natural in a free-field
theory, is so appropriate in Liouville theory.

     If one wants to make use of results from free-field theory in order
to study the dynamics of Liouville theory, it is more appropriate to use the
canonical transformation as implemented by intertwining operators, as we
have discussed in section two.  In contrast to the operator-state map,
which is a {\it time}-asymptotic construction that aims to obtain Liouville
eigenfunctions from an initial plane-wave form of the wavefunction, the
intertwining-operator construction may be thought of as deriving the
Liouville wavefunctions from {\it spatially}-asymptotic limits. One may
view the effect of the canonical transformation this way because the exact
wavefunctions $\psi_p(q)$ asymptotically approach linear combinations of
free-field exponentials $e^{\pm\im pq}$ as $q\rightarrow-\infty$. As we have
seen, this spatially-asymptotic identification holds only up to Weyl-group
transformations: the two-to-one nature of the canonical transformation is
expressed by the fact that both $e^{\im p\tilde q}$ and $e^{-\im p\tilde
q}$ are transformed to $\psi_p(q)$.

\bigskip
\noindent {\bf 4. Canonical transformations for Toda gravity}
\medskip

     For the minimal (two-scalar) non-critical $W_3$ string, we derived the
allowable forms of the potentials in Eqs (1.8). It will now be convenient to
make  a field redefinition in order to bring the resulting minisuperspace
Toda Lagrangian to  a standard form. Accordingly, we set
$$\eqalignno{
\varphi_1(\tau,\sigma)&=-{4Q_1\over7}q_1(\tau) +
\varphi^{\rm osc}_1(\tau,\sigma)&(4.1a)\cr
\varphi_2(\tau,\sigma)&=-{4Q_2\over7}(2q_2(\tau)-q_1(\tau)) +
\varphi^{\rm osc}_2(\tau,\sigma),&(4.1b)\cr
\oint d\sigma\,\varphi^{\rm osc}_{1}&=\oint d\sigma\,\varphi^{\rm
osc}_{2}=0.&(4.1c)\cr}
$$ The action for the minisuperspace modes $q_1$ and $q_2$ in this
basis becomes, again denoting the time parameter for the
quantum-mechanical system by $t$,
$$
I_{\rm T}=\int dt\,(\dot q_1^2+\dot q_2^2-\dot q_1\dot q_2 -e^{2q_1-q_2}
-e^{2q_2-q_1}),\eqno(4.2)
$$
which can be recognised as the action for Toda mechanics in the
Chevalley basis for $A_2\equiv SU(3)$. The Hamiltonian following from (4.2)
is then
$$
H_{\rm
T}=\ft13(p_1^2+p_2^2+p_1p_2)+e^{2q_1-q_2}+e^{2q_2-q_1},\eqno(4.3)
$$
giving equations of motion
$$
\eqalignno{ p_1=2\dot q_1-\dot q_2\qquad&\qquad p_2=2\dot q_2-\dot
q_1&(4.4a)\cr
\ddot q_1=-e^{2q_1-q_2}\qquad&\qquad \ddot
q_2=-e^{2q_2-q_1}.&(4.4b)\cr}
$$

\bigskip
\noindent{\it 4.1 Classical Toda Mechanics}
\medskip

     Proceeding in analogy to our Liouville discussion of section 2,
we first  seek solutions to the classical Toda mechanics equations
(4.4), following the  style of Ref.\ [35], starting from the
ansatz
$$
e^{-q_1}=\sum_{i=1}^3 f_i\,e^{\mu_it}\qquad\qquad
e^{-q_2}=\sum_{i=1}^3 g_i\,e^{-\mu_it},\eqno(4.1.1)
$$
where $\mu_i$, $f_i$ and $g_i$ are constants. Substituting into
(4.4$b$), we  find that one must require
$$
\eqalignno{
\mu_1+\mu_2+\mu_3=0&&(4.1.2a)\cr
f_1f_2f_3(\mu_1-\mu_2)^2(\mu_2-\mu_3)^2(\mu_3-\mu_1)^2=1&&(4.1.2b)\cr
g_1=f_2f_3(\mu_2-\mu_3)^2&,\quad {\rm and\ cyclically.}&(4.1.2c)\cr}
$$
Solutions of this form, which have four free parameters after
taking Eqs  (4.1.2) into account, are in fact general solutions to the
classical equations of motion (4.4).

     It is convenient to write $\mu_1t=\tilde q_1$ and $\mu_2t=\tilde
q_2$ and to reparametrise the general solution (4.1.1) subject to
(4.1.2), as
$f_1=g_1=[\tilde p_1(\tilde p_1-\tilde p_2)]^{-1}$, $f_2=g_2=[\tilde
p_2(\tilde p_1-\tilde p_2)]^{-1}$, $f_3=g_3=[\tilde p_1\tilde
p_2]^{-1}$. The solution to (4.4) then takes the form
$$
\eqalignno{ e^{-q_1}&={1\over \tilde p_1(\tilde p_1-\tilde
p_2)}e^{\tilde q_1} + {1\over
\tilde p_2(\tilde p_1-\tilde p_2)}e^{\tilde q_2}+{1\over \tilde
p_1\tilde  p_2}e^{-\tilde q_1-\tilde q_2}&(4.1.3a)\cr
e^{-q_2}&={1\over \tilde p_1(\tilde p_1-\tilde p_2)}e^{-\tilde q_1} +
{1\over
\tilde p_2(\tilde p_1-\tilde p_2)}e^{-\tilde q_2}+{1\over \tilde
p_1\tilde  p_2}e^{\tilde q_1+\tilde q_2}&(4.1.3b)\cr
(2p_1+p_2)e^{-q_1}&=-{(2\tilde p_1-\tilde p_2)\over \tilde p_1(\tilde
p_1-\tilde p_2)}e^{\tilde q_1} - {(2\tilde p_2-\tilde p_1)\over \tilde
p_2(\tilde p_1-\tilde p_2)}e^{\tilde q_2}+{(\tilde p_1+\tilde
p_2)\over \tilde p_1\tilde p_2}e^{-\tilde q_1-\tilde q_2}&(4.1.3c)\cr
(2p_2+p_1)e^{-q_2}&={(2\tilde p_1-\tilde p_2)\over \tilde p_1(\tilde
p_1-\tilde p_2)}e^{-\tilde q_1} + {(2\tilde p_2-\tilde p_1)\over \tilde
p_2(\tilde p_1-\tilde p_2)}e^{-\tilde q_2}-{(\tilde p_1+\tilde
p_2)\over \tilde p_1\tilde p_2}e^{\tilde q_1+\tilde q_2}.&(4.1.3d)\cr}
$$

     The solution (4.1.3) may be interpreted as a canonical
transformation
$F$ from the interacting Toda-theory variables $(q_1,q_2;p_1,p_2)$ to
free  variables $(\tilde q_1,\tilde q_2;\tilde p_1,\tilde p_2)$. One
may verify  directly that this transformation is indeed canonical. The
Toda Hamiltonian (4.3) maps into the free
Hamiltonian\footnote{$^{\ast}$}{\tenfoot Note that one could bring the the
free Hamiltonian into the same form as the kinetic part of the interacting
Hamiltonian (4.3), {\it e.g.}\ by tacking onto (4.1.3) an additional
canonical transformation $(p_1,q_1)\mapsto(-\tilde p_1,-\tilde q_1)$. We
shall not do so here, but shall need to include such an additional
transformation later on for comparison with results derived in the
literature for the free-field $W_3$ string, without the interaction
potentials (1.8).}
$$
\tilde H_{\rm T}=\ft13(\tilde p_1^2+\tilde p_2^2-\tilde p_1\tilde
p_2).\eqno(4.1.4)
$$
Note that this implies that the free momenta $\tilde p_1$ and
$\tilde p_2$ are related to the free velocities by
$$
\tilde p_1=2\dot{\tilde q}_1 + \dot{\tilde q}_2\qquad\qquad\tilde
p_2=2\dot{\tilde q}_2 + \dot{\tilde q}_1.\eqno(4.1.5)
$$

     The exact classical integrability of the Toda system has another
aspect, namely the existence of a second independent constant of the motion.
This is of course just the mini\-superspace limit of the spin-3 current
(1.6) of the $W_3$ string, after taking into account appropriate shifts of
the momenta by background charges as in (2.3). For the interacting theory
(4.2), this second invariant is
$$\eqalign{
W_{\rm T}&=\ft19p_1^3+\ft16p_1^2p_2-\ft16p_1p_2^2-\ft19p_2^3 +\ft12
(2p_2+p_1) e^{2q_1-q_2}-\ft12 (2p_1+p_2) e^{2q_2-q_1}\cr
&=\ft1{18}(2p_1+p_2)(2p_2+p_1)(p_1-p_2) +\ft12 (2p_2+p_1)
e^{2q_1-q_2}-\ft12 (2p_1+p_2) e^{2q_2-q_1}.\cr}
\eqno(4.1.6)
$$
Upon use of the canonical transformation $F$ (4.1.3), this becomes
a purely cubic invariant of the free theory,
$$
\eqalign{
\tilde W_{\rm T}&=-\ft19\tilde p_1^3+\ft16\tilde p_1^2\tilde p_2 +
\ft16\tilde p_1\tilde p_2^2-\ft19\tilde p_2^3\cr &=\ft1{18}(2\tilde
p_1-\tilde p_2)(2\tilde p_2-\tilde p_1)(\tilde p_1+ \tilde
p_2). \cr}\eqno(4.1.7)
$$

     The two invariants $\tilde H_{\rm T}$ and $\tilde W_{\rm T}$
determine the discrete symmetries of the free theory obtained from the
Toda theory by our canonical transformation (4.1.3). These
discrete symmetries in turn determine the multiplet of forms of the canonical
transformation, in analogy to (2.1.3), (2.1.7). Requiring that (4.1.4) and
(4.1.7) be symmetric determines the following group of
transformations. Firstly, the free theory has a threefold symmetry under
rotations of the free momenta $\tilde p_1$ and $\tilde p_2$ accompanied by
corresponding rotations of the free variables $\tilde  q_1$ and
$\tilde q_2$; these are generated by
$$
M:\qquad (\tilde q_1,\tilde q_2;\tilde p_1,\tilde p_2)\rightarrow (-\tilde
q_1-\tilde  q_2,\tilde q_1;-\tilde p_2,\tilde p_1-\tilde p_2).\eqno(4.1.8)
$$
The full threefold symmetry is obtained by acting with $M^0$, $M^1$ or $M^2$.
In addition, there is a twofold reflection symmetry,
$$
R:\qquad (\tilde q_1,\tilde q_2;\tilde p_1,\tilde p_2)\rightarrow (\tilde
q_2,\tilde q_1;\tilde p_2,\tilde p_1),\eqno(4.1.9)
$$
so that the full discrete symmetry group for the free theory as determined
by the invariants (4.1.4), (4.1.7) is $S_3$, the permutation group of three
objects. As foreshadowed by our Liouville discussion in section two, this is
precisely the Weyl group of $A_2\equiv SU(3)$, the underlying algebra of the
Toda theory.

     Another important analogue of the Liouville case is the fact that the
classical canonical transformation $F$ (4.1.3) has only a limited domain of
applicability in the ($\tilde p_1$,$\tilde p_2$) plane, since the left-hand
sides of (4.1.3$a$,$b$) are always positive. Other domains in the
free-momentum plane are obtained by applying the Weyl-group
transformations (4.1.8), (4.1.9) to the free-theory variables in $F$,
generating a corresponding Weyl multiplet of forms of the canonical
transformation. We start from a reference region
in the free-momentum plane, in which (4.1.3$a$,$b$) are clearly applicable:
$$
\tilde p_2>0;\quad\tilde p_1>\tilde p_2,\eqno(4.1.10)
$$
which is in fact the principal Weyl chamber of $SU(3)$ in the Chevalley
basis.
Note that in this basis, the Weyl chambers occupy $45^\circ$ and $90^\circ$
wedges rather than the uniform $60^\circ$ wedges of an orthonormal basis.
\bigskip
\centerline{\epsfbox{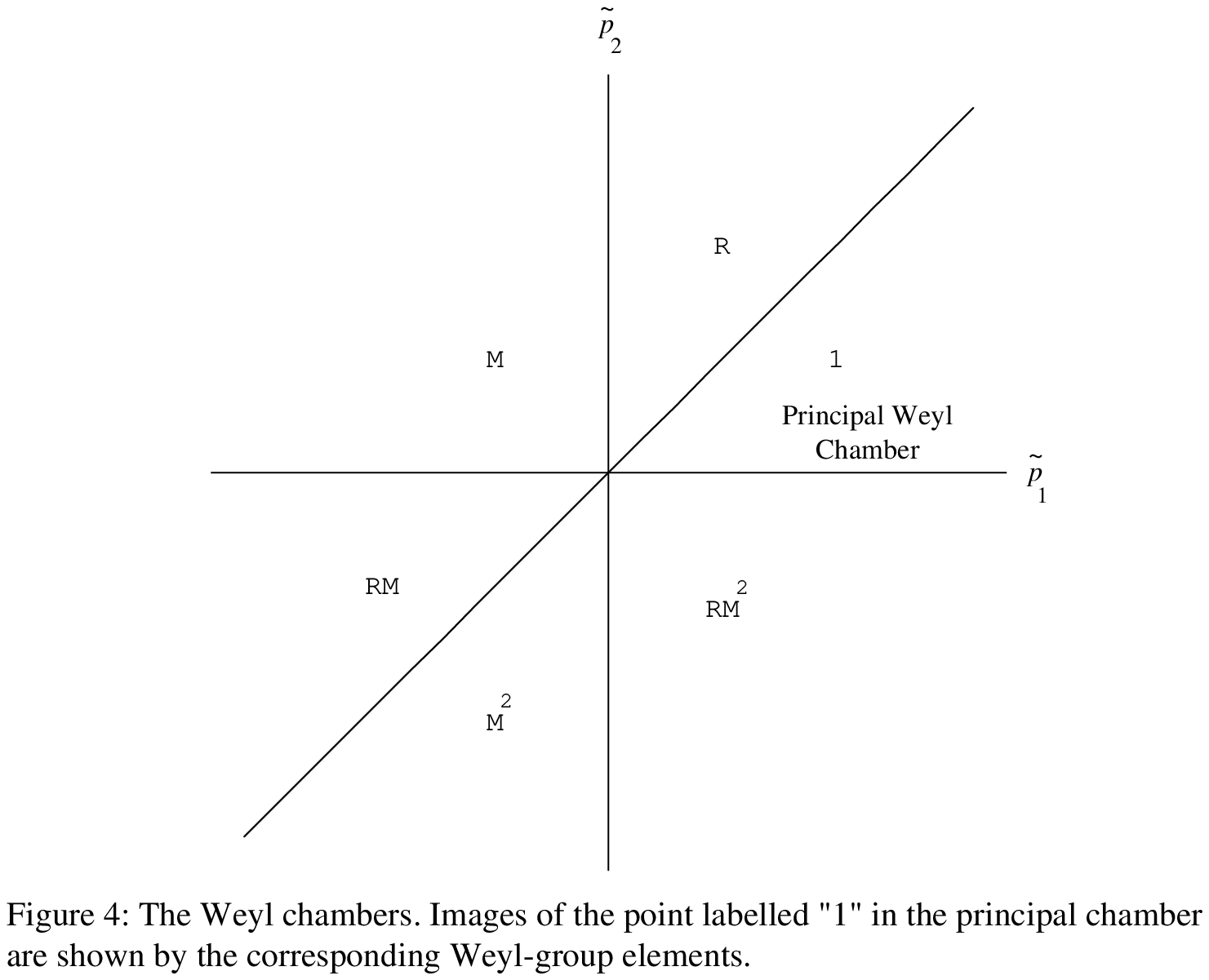}}

     It is easy to check that, by acting with $M$ zero, one or two times and
with $R$ zero or one time, one generates six non-overlapping regions
whose union covers the entire free-momentum plane, as shown in Figure 4.
In order to link these domains with the corresponding forms of the canonical
transformation, it is necessary to recognise another feature of the
transformation: there is in fact only a {\it triplet} of distinctly different
maps obtained from $F$ by application of the Weyl group, because (4.1.3)
is invariant under the order-two transformation
$$
Z:\qquad (\tilde q_1,\tilde q_2;\tilde p_1,\tilde p_2)\rightarrow (-\tilde
q_1-\tilde  q_2,\tilde q_2;-\tilde p_1,-\tilde p_1+\tilde p_2),\eqno(4.1.11)
$$
which is the Weyl group element $RM$. Acting with this transformation on the
reference region (4.1.10) generates the region
$$
\tilde p_2<0;\quad\tilde p_1<\tilde p_2,\eqno(4.1.12)
$$
obtained from the former by inversion through the origin of the
free-momentum plane. Eqs (4.1.3$a$,$b$) are once again clearly applicable
in the region (4.1.12). Owing to the invariance (4.1.11), we have, even for
the starting transformation $F$, a situation similar to that of (2.1.7) in
the Liouville case: the inverse of (4.1.3) has a twofold branch structure,
because one has to choose a set of free momenta either from the region
(4.1.10) or from the region (4.1.12). Thus, the fundamental reference
region within which (4.1.3) has a uniquely-defined inverse may be taken to
be the principal Weyl chamber (4.1.10). Upon application of the rotations
$M^0$, $M^1$, $M^2$ to the  pair of regions (4.1.10, 4.1.12), one then
completely covers the free-momentum plane in a non-overlapping fashion; the
three images of this pair are in one-to-one correspondence with the
distinct transformations $F$, $MF$, $M^2F$.

     In sum, application of the Weyl group $S_3$ to our starting canonical
transformation $F$ (4.1.3) and to its principal Weyl-chamber region (4.1.10)
produce three distinct forms of the transformation and six distinct Weyl
chambers. Considering the union of these regions together with their
respective forms of the transformation as a single canonical-transformation
map, we see that one has a six-fold branch structure in the $A_2$ Toda
case, giving a six-to-one relation between free-field solutions and the
interacting Toda-theory solutions.

     As earlier in the Liouville case, it is instructive to examine the
asymptotic relationship between the free-field momenta and the interacting
Toda-theory momenta for solutions to the classical equations of motion
(4.4). Consider the transformation $F$ (4.1.3) for free momenta in the
principal Weyl chamber (4.1.10) in the limit as $t\rightarrow-\infty$. In
(4.1.3$a$), the dominant term on the right-hand side in this limit is
$e^{-(\tilde q_1+\tilde q_2)}=e^{-\ft13(\tilde p_1+\tilde p_2)t}$, using
relation (4.1.5) between the free momenta and velocities. Thus, in this
limit, we have $e^{-q_1}\rightarrow(\tilde p_1\tilde p_2)^{-1}e^{-(\tilde
q_1+\tilde q_2)}$. Similarly, in (4.1.3$b$), the dominant term in the
$t\rightarrow-\infty$ limit is $e^{-\tilde q_1} = e^{-\ft13(2\tilde
p_1-\tilde p_2)t}$, so we obtain
$e^{-q_2}\rightarrow [\tilde p_1(\tilde p_1-\tilde p_2)]^{-1}e^{-\tilde
q_1}$. Then, from (4.1.3$c$,$d$) we obtain the asymptotic relations between
Toda and free momenta: $p_1\rightarrow\tilde p_2$; $p_2\rightarrow\tilde
p_1-\tilde p_2$. Although the details of the asymptotic relations change as
one goes through the six different Weyl chambers, the pattern that we found
in our Liouville discussion is repeated here: Weyl-group images of the
starting free-field momenta ($\tilde p_1$, $\tilde p_2$) all describe one
and the same Toda solution. Thus, we may now concentrate just on the
transformations (4.1.3) for the Weyl chamber (4.1.10).

     The region of classically-realisable asymptotic Toda momenta for
$t\rightarrow-\infty$ following from the above discussion is the whole
first quadrant of the Toda-momentum plane:
$p_1>0$, $p_2>0$. The Toda velocities obtained from (4.1.5) are $\dot
q_1=\ft13(2p_1+p_2)$, $\dot q_2=\ft13(2p_2+p_1)$. Thus, the
classically-realisable velocities for $t\rightarrow-\infty$ must lie in the
wedge $2\dot q_2-\dot q_1>0$, $2\dot q_1-\dot q_2>0$. This wedge
corresponds precisely to the asymptotic directions of approach allowed by
the ``valley'' in the Toda potential, as shown in Figure 5. Unlike the
Liouville case, where by dimensionality the motion is constrained to aim
straight at the ``mountain'', the Toda case in one dimension higher affords
a trajectory of motion aimed at the ``ravine,'' or mildest ascent, in the
direction opposite to the valley.
\bigskip
\centerline{\epsfbox{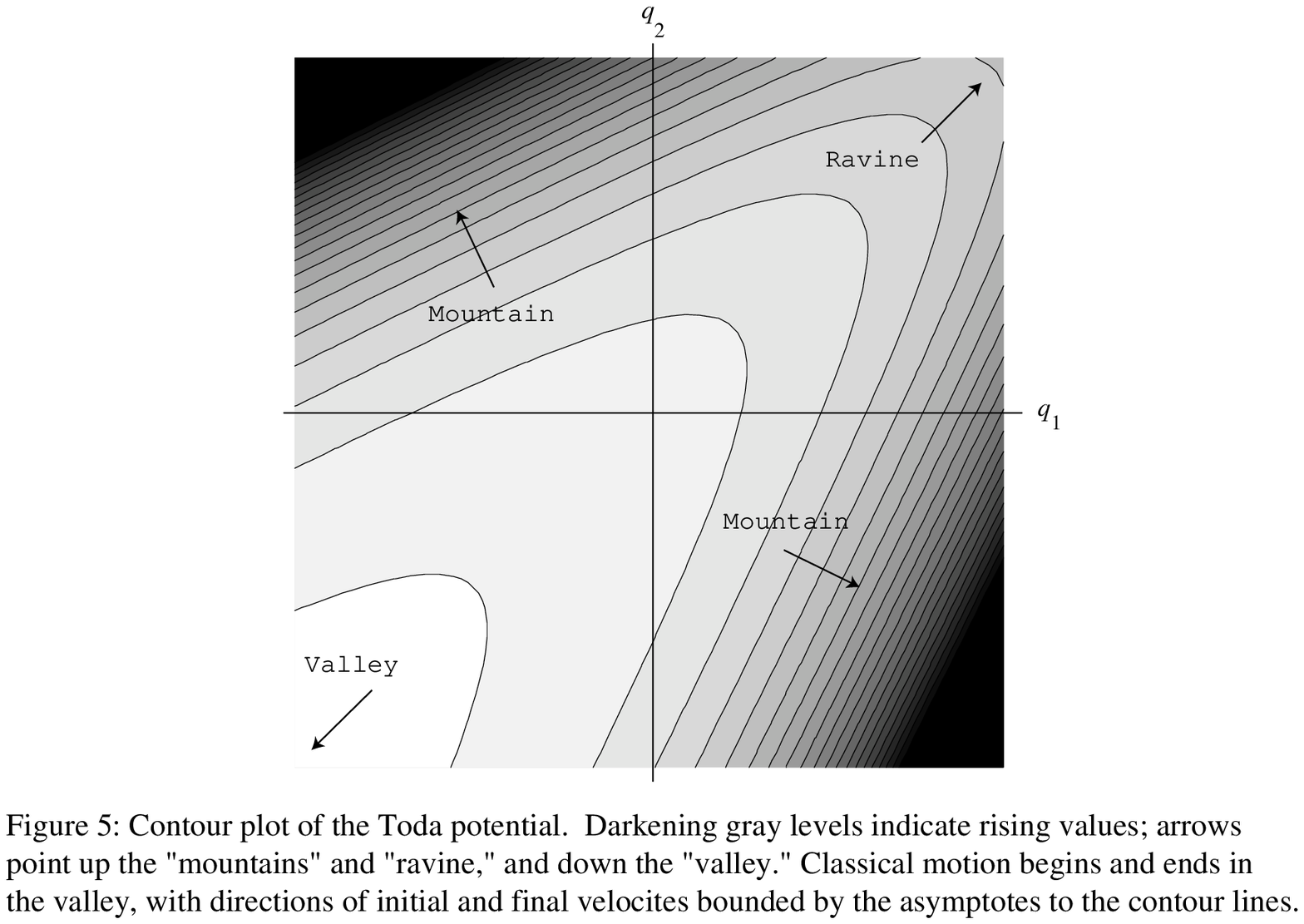}}

     One may make an analogous asymptotic analysis of the motion for
$t\rightarrow+\infty$, considering again only the free solutions
corresponding to (4.1.3), (4.1.10). In this case, for
$e^{-q_1}$ the dominant term is $e^{\tilde q_1}$ and for $e^{-q_2}$ the
dominant term is $e^{\tilde q_1+\tilde q_2}$. Thus, one obtains for
$t\rightarrow+\infty$ the limits $e^{-q_1}\rightarrow[\tilde p_1(\tilde
p_1-\tilde p_2)]^{-1}e^{\tilde q_1}$, $e^{-\tilde q_2}\rightarrow(\tilde
p_1\tilde p_2)^{-1}e^{\tilde q_1+\tilde q_2}$, and then $p_1\rightarrow
-\tilde p_1+\tilde p_2$, $p_2\rightarrow-\tilde p_2$. With respect to the
$t\rightarrow-\infty$ asymptotic momenta, the
$t\rightarrow+\infty$ limits differ precisely by an application of the Weyl
group element $Z=RM$ (4.1.11). Thus, as we found earlier in the Liouville
case, the classical asymptotic interacting-theory momenta at
$t\rightarrow\pm\infty$ differ by a Weyl-group transformation of the free
momenta. Thus, in the Toda theory as well in the Liouville on, asymptotic
values of interacting-theory momenta are conserved {\it modulo} the Weyl
group.

\np
\noindent{\it 4.2 Quantum Toda Mechanics}
\medskip

     The quantum generalisation of the canonical transformation (4.1.3) once
again maps the interacting-theory Hamiltonian (4.3) into the free-theory
Hamiltonian (4.1.4), but this time care needs to be taken of quantum operator
orderings. We find that the following sequence of elementary canonical
transformations implements the free-field map:
\crampest
$$\matrix{\rlap{[T1]}\hskip.5cm&{e^{\ft\pi2(p_1+p_2)}\over\Gamma(1-\im(p_1+p_2))}:
&\rlap{$\left\{
\matrix{e^{q_1}\mapsto -e^{q_1}(p_1+p_2),\cr e^{q_2}\mapsto
-e^{q_2}(p_1+p_2),\cr}\right.$}\hskip 4.6cm&\rlap{$\matrix{p_1\mapsto p_1\cr
p_2\mapsto p_2\cr}$}\hskip 5.75cm\cr\strut\cr
\rlap{[T2]}\hskip.5cm&{\cal P}_{(\ln q_1,\ln q_2)}:&\rlap{$\left\{
\matrix{q_1\mapsto\ln q_1,\cr q_2\mapsto\ln q_2,\cr}\right.$}
\hskip 4.6cm&\rlap{$\matrix{ p_1\mapsto q_1p_1\cr p_2\mapsto q_2p_2\cr}$}
\hskip 5.75cm\cr\strut\cr
\rlap{[T3]}\hskip.5cm&q_1^{-1}q_2^{-2}:&\rlap{$\left\{
\matrix{q_1\mapsto q_1,\cr q_2\mapsto q_2,\cr}\right.$}\hskip 4.6cm&
\rlap{$\matrix{p_1\mapsto p_1-{\im\over q_1}\cr  p_2\mapsto p_2-{\im\over
q_2}\cr}$}\hskip 5.75cm\cr\strut\cr
\rlap{[T4]}\hskip.5cm &\exp\left(-\im({q_1^2\over q_2} + {q_2^2\over
q_1})\right):&\rlap{$\left\{
\matrix{q_1\mapsto q_1 ,\cr q_2\mapsto q_2,\cr}\right.$}
\hskip 4.6cm&\rlap{$\matrix{p_1\mapsto p_1-{q_2^2\over q_1^2}+{2q_1\over q_2}
\cr p_2\mapsto p_2-{q_1^2\over q_2^2}+{2q_2\over q_1}\cr}$}
\hskip 5.75cm\cr\strut\cr
\rlap{[T5]}\hskip.5cm&{\cal I}_1{\cal I}_2:&\rlap{$\left\{
\matrix{q_1\mapsto p_1,\cr q_2\mapsto p_2,\cr}\right.$}
\hskip 4.6cm&\rlap{$\matrix{p_1\mapsto -q_1\cr p_2\mapsto -q_2\cr}$}
\hskip 5.75cm\cr\strut\cr
\rlap{[T6]}\hskip.5cm&\rlap{${\cal P}_{(q_1-q_2+{1\over q_1q_2},{1\over
q_1}-{1\over q_2}+q_1q_2)}:$}\hskip4.25cm&\rlap{$\left\{
\matrix{q_1\mapsto q_1'=\phantom{\rlap{${1\over\det{\partial q_i'\over
\partial q_j}}$}}q_1-q_2+{1\over q_1q_2},\cr q_2\mapsto
q_2'=\phantom{\rlap{${1\over\det{\partial q_i'\over\partial q_j}}$}}{1\over
q_1}-{1\over q_2}+q_1q_2,\cr}\right.$}
\hskip 4.6cm&\rlap{$\matrix{p_1\mapsto p_1'={1\over\det{\partial
q_i'\over\partial q_j}}\left({\partial q_2'\over\partial q_2}p_1- {\partial
q_2'\over\partial q_1}p_2\right)\cr p_2\mapsto
p_2'={1\over\det{\partial q_i'\over\partial q_j}}\left({\partial
q_1'\over\partial q_1}p_2-{\partial q_1'\over\partial q_2}p_1\right)\cr}$}
\hskip 5.75cm\cr\strut\cr
\rlap{[T7]}\hskip.5cm&{\cal P}_{(e^{q_1},e^{q_2})}:&\rlap{$\left\{
\matrix{q_1\mapsto e^{q_1},\cr q_2\mapsto e^{q_2},\cr}\right.$}
\hskip 4.6cm&\rlap{$\matrix{p_1\mapsto e^{-q_1}p_1\ \cr p_2\mapsto
e^{-q_2}p_2.\cr}$}\hskip 5.75cm\cr}
$$
\uncramp

     Among the transformations composing this free-field map, we have a
type not yet encountered in this paper, the ``similarity'' transformations
embodied in [T1,T3,T4] (although, strictly speaking, the inverse-momentum
transformations [L3] are also of this type). The coordinate similarity
transformations [T3,T4], of the form ($p_i\mapsto p_i-f,_i(q_j)$,
$q_i\mapsto q_i$) and are generated by $e^{\im f(q_j)}$, transforming
wavefunctions
$\Psi(q_j)$ into $e^{\im f(q_j)}\Psi(q_j)$ [19,21]. Momentum versions such
as [T1], of the form  ($q_i\mapsto q_i+f,_i(p_j)$, $p_i\mapsto p_i$), are
generated by
$e^{\im f(p_j)}={\cal I} e^{\im f(q_j)} {\cal I}^{-1}$. The Gamma function in
the generator of [T1] may be given an integral representation
$$
\Gamma(1-\im p_1-\im p_2)=\int_0^\infty du\, u^{-\im p_1-\im p_2}e^{-u},
\eqno(4.2.1)
$$ in which the momentum operators in the exponent simply generate
translations:
$$
u^{-\im p_1-\im p_2}=\exp\Big(-\im(p_1+p_2)\ln u\Big).\eqno(4.2.2)
$$

     Following the mapping of the Toda Hamiltonian (4.3) as the quantum
canonical transformation [T1 -- T7] unfolds, we have the sequence of
stages
$$
\eqalign{ 3 H_{\rm T}&=p_1^2+p_2^2+p_1p_2+3e^{2q_1-q_2}+3e^{2q_2-q_1}\cr
[{\rm T1}]\qquad\qquad&\mapsto p_1^2+p_2^2+p_1p_2-3(e^{2q_1-q_2}+
e^{2q_2-q_1})(p_1+p_2)\cr
[{\rm T2}]\qquad\qquad&\mapsto
(q_1p_1)^2+(q_2p_2)^2+q_1p_1q_2p_2- 3\left({q_1^2\over q_2}+{q_2^2\over
q_1}\right)(q_1p_1+q_2p_2)\cr
[{\rm T3}]\qquad\qquad&\mapsto
(p_1q_1)^2+(p_2q_2)^2+p_1q_1p_2q_2- 3\left({q_1^2\over q_2}+{q_2^2\over
q_1}\right)(p_1q_1+p_2q_2)\cr
[{\rm T4}]\qquad\qquad&\mapsto
(p_1q_1)^2+(p_2q_2)^2+p_1q_1p_2q_2 -9q_1q_2-3p_2q_1^2-3p_1q_2^2\cr
[{\rm T5}]\qquad\qquad&\mapsto (q_1p_1)^2+(q_2p_2)^2+q_1p_1q_2p_2
-9p_1p_2+3q_2p_1^2+3q_1p_2^2\cr
[{\rm T6}]\qquad\qquad&\mapsto
(q_1p_1)^2+(q_2p_2)^2-q_1p_1q_2p_2\cr
[{\rm T7}]\qquad\qquad&\mapsto
p_1^2+p_2^2-p_1p_2=3\tilde H_{\rm T}.\cr}\eqno(4.2.3)
$$

     The mapping from free-field wavefunctions to interacting Toda
wavefunctions is then given by the inverse transformation generator
\crampest
$$\eqalign{
C^{-1}= e^{-\ft\pi2(p_1+p_2)}\Gamma(1-\im p_1-\im p_2)&{\cal
P}_{(e^{q_1},e^{q_2})} q_1q_2\exp\left(\im({q_1^2\over q_2} + {q_2^2\over
q_1})\right)\cdot\cr
&{\cal I}_1^{-1}{\cal I}_2^{-1} {\cal
P}^{-1}_{(q_1-q_2+{1\over q_1q_2},{1\over q_1}-{1\over q_2}+q_1q_2)}{\cal
P}_{(\ln q_1,\ln q_2)};\cr}\eqno(4.2.4)
$$
\uncramp
so that
$$
\Psi_{k_1,k_2}(q_1,q_2)=N_{k_1k_2}C^{-1} e^{ik_1 q_1+ik_2 q_2},\eqno(4.2.5)
$$
where $N_{k_1k_2}$ is a normalization constant. To obtain a well-defined
integral representation of the Toda wavefunction,  the contours and ranges of
integration for the Fourier transforms must be  specified. These will be
chosen so as to give a wavefunction with the physically correct asymptotic
behaviour. In order to have a normalizable Toda wavefunction, the
wavefunction must vanish asymptotically in the region under the  potential,
{\it i.e.}\ in the region where $2q_1-q_2>0,\ 2q_2-q_1>0$.

To evaluate the wavefunction, focus first on the portion of the
transformation (4.2.4) from the initial point transformation to the
inverse Fourier transforms, {\it i.e.}\
$$ {\cal I}_1^{-1}{\cal I}_2^{-1} {\cal P}^{-1}_{(q_1-q_2+{1\over
q_1q_2},{1\over q_1}-{1\over q_2}+q_1q_2)}{\cal P}_{(\ln q_1,\ln q_2)}e^{\im
k_1 q_1+\im k_2 q_2}.\eqno(4.2.6)
$$
Formally, this is
$$
{1\over2\pi}\int dz_1 \int dz_2 e^{-\im q_1 z_1} e^{-\im q_2 z_2} {\cal
P}^{-1}_{(y_1-y_2+{1\over y_1y_2},{1\over y_1}-{1\over y_2}+y_1y_2)}y_1^{\im
k_1}y_2^{\im k_2}.\eqno(4.2.7)
$$
We shall determine the region of integration shortly. The
inverse point transformation ${\rm [T6]}^{-1}$ implements the change of
variables $z_1=y_1-y_2+{1\over y_1y_2}$, $z_2={1\over y_1}-{1\over
y_2}+y_1y_2$; consequently, one obtains the integral
\def\jac{[{\rm jac}]}
$$
{1\over2\pi}\int dy_1 \int dy_2\, \jac\, y_1^{\im k_1} y_2^{\im k_2}
e^{-\im q_1(y_1-y_2+{1\over y_1y_2})}e^{-\im q_2({1\over y_1}- {1\over
y_2}+y_1y_2)}.\eqno(4.2.8)
$$
where
$$
\jac={-1\over y_1 y_2}(y_1+y_2)(-y_2+{1\over y_1^2}) (y_1+{1\over
y_2^2}).\eqno(4.2.9)
$$

     Applying the remaining transformations subsequent to the inverse Fourier
transforms ${\rm [T5]}^{-1}$ in (4.2.4), and acting with the translation
operator inside the Gamma function, the Toda wavefunction is found to be
$$
\eqalign{
\Psi_{k_1,k_2}(q_1,q_2)&={N_{k_1k_2}\over2\pi}e^{\im\pi}\int_0^\infty du\,
e^{q_1+q_2}u^{-2}e^{-u-(e^{2q_1-q_2}+e^{2q_2-q_1})u^{-1}}\times\cr
&\qquad\int dy_1
\int dy_2 \,\jac\, y_1^{\im k_1} y_2^{\im k_2}  e^{e^{q_1}(y_1-y_2+{1\over
y_1y_2})u^{-1}}e^{e^{q_2}({1\over y_1}-{1\over y_2}+y_1y_2)u^{-1}}.\cr}
\eqno(4.2.10)
$$
The $y_1$, $y_2$ integrals are
convergent if the ranges of integration are taken to run from $-\infty$ to
$0$ and from $0$ to $\infty$, respectively. The $u$ integral is then seen to
be convergent. These ranges are also the ones that give vanishing
boundary contributions when integrating by parts in verifying that the
generalised inverse Fourier transforms ${\cal I}_1^{-1}{\cal I}_2^{-1}$
correspond correctly to the inverse of the operator transformations [T5],
analogously to our discussion in section two for the Liouville case.

     In order to make the convergence of the integrals more explicit, we
redefine $y_1\mapsto -y_1$, obtaining the manifestly convergent expression
$$
\eqalign{
\Psi_{k_1,k_2}(q_1,q_2)&={N_{k_1k_2}\over2\pi}e^{\pi k_1}\int_0^\infty
du\,  e^{q_1+q_2}u^{-2}e^{-u-(e^{2q_1-q_2}+e^{2q_2-q_1})u^{-1}}\times\cr
&\qquad\int_0^\infty dy_1 \int_0^{\infty} dy_2
\,\jac'\, y_1^{\im k_1} y_2^{\im k_2}  e^{ -e^{q_1}(y_1+y_2+{1\over
y_1y_2})u^{-1}} e^{ -e^{q_2}({1\over y_1}+{1\over y_2}+y_1y_2)u^{-1}},\cr
}\eqno(4.2.11)
$$  where
$$
\jac'={1\over y_1 y_2}(y_1-y_2) (y_2-{1\over y_1^2}) (y_1-{1\over
y_2^2}).\eqno(4.2.12)
$$
The wavefunction is now seen to vanish strongly under the potential owing to
the presence of the factor $\exp(-(e^{2q_1-q_2}+e^{2q_2-q_1})u^{-1})$.

     The Weyl-group symmetry of the wavefunction may now be seen directly in
(4.2.11) by changing the variables of integration, again analogously
to our earlier Liouville discussion. The $M$ transformation of
the Weyl group corresponds to the order-three change of variables
\def\Mto{\buildrel M\over\longrightarrow}
\crampest
$$
\matrix{y_1&\Mto &y_2&\Mto &{1\over y_1y_2}\cr
y_2&\Mto &{1\over y_1y_2}&\Mto &y_1\cr}\qquad \Leftrightarrow\qquad
\matrix{k_1&\Mto &-k_2&\Mto &k_2-k_1\cr
k_2&\Mto &k_1-k_2&\Mto &-k_1\cr},\eqno(4.2.13)
$$
\uncramp
while the $R$ transformation corresponds to the order-two change of
variables
\def\Rto{\buildrel R\over\longrightarrow}
\crampest
$$
\matrix{y_1&\Rto &y_2\cr y_2&\Rto &y_1\cr}
\qquad \Leftrightarrow\qquad
\matrix{k_1&\Rto &k_2\cr k_2&\Rto &k_1\cr}.\eqno(4.2.14)
$$
\uncramp
Thus, the only changes to the wavefunction (4.2.11) that arise upon making a
Weyl-group transformation are the inessential ones occurring in the prefactor
$e^{\pi k_1}$; even these will in the end be cancelled after
normalization by compensating changes in $N_{k_1,k_2}$, as we saw in the
Liouville case (2.2.9). Thus, the quantum Toda theory possesses the full
Weyl-group symmetry that we found for Toda theory at the classical level.

\bigskip
\noindent {\bf 5. Conclusion}
\bigskip

     In this paper, we have made explicit constructions of the wavefunctions
of Liouville and Toda gravities, using the technique of intertwining
operators to realise canonical transformations between the free and
interacting theories. These detailed results show distinctly the very
important r\^ole played by the Weyl group of the underlying $A_N$ Lie algebra
in each case. At the classical level, this symmetry is seen through the
presence of a multiplet of forms of the free-field canonical transformation
map, each form being physically appropriate for a different domain in the
space of free-field momenta. At the quantum level, the canonical
transformations of the fundamental conjugate operator pairs
$(p_i,q_i)$ must be constructed with careful attention to operator ordering.

     In analysing the Weyl-group symmetries of the free-field maps, it
emerges in both the Liouville and Toda cases that the most Weyl-symmetric
form of the quantum canonical transformation does not itself have a direct
limit to a sensible classical transformation. We saw this in the complex
character of the manifestly Weyl-symmetric transformation (2.2.15) for the
operator pair $(p,q)$ in the Liouville case. The Weyl-symmetric
intertwining-operator map (4.2.11) in the Toda case gives rise to a
similarly manifestly symmetric transformation of the operator pairs
$(p_i,q_i)$, and, once again, the classical limit of this transformation can
be seen to admit no real classical solutions. The manifestly
Weyl-symmetric forms of these quantum transformations also generate the most
manifestly-convergent forms of the quantum wavefunctions. Moreover, these
manifestly-symmetric forms are also the ones that manifestly satisfy the
physically-required boundary condition of falling away to zero in the region
under the Liouville/Toda potentials. It appears that, in requiring manifest
Weyl symmetry, as well as in imposing the physical boundary conditions, one
is focusing on behaviour in the region under the potential, where motion is
classically excluded. In both the Liouville and Toda cases, there exist
other forms of the quantum transformations, obtained by analytic
continuation, that do limit to classically-sensible forms [L1--L4, T1--T7].
In these classically-sensible forms of the quantum transformations, the
Weyl symmetry is harder to spot, and also the integral
representations for the wavefunctions lie just at the limits of convergence
and hence need to be regularized. Such behaviour is of course familiar from
general comparisons between Euclidean and Minkowskian approaches to quantum
theory, but the involvement of the Weyl-group symmetry in the present cases
is striking.

     Another striking result of our investigations is the precise nature of
the operator-state map, which figures in many discussions of non-critical
string theory. It is important to emphasise that the operator-state map is
quite distinct from the true canonical transformation between free and
interacting fields that we have derived using intertwining operators. Some
aspects of the operator-state map, such as the occurrence of Seiberg
bounds, are specific properties of this map and do not particularly reveal
features of the underlying dynamics. Use of the operator-state map is
motivated by the success of conformal field theory in formulating
free-field string theory in terms of vertex operators. It is not clear to
us how successfully the vertex-operator approach can be carried over to the
interacting Liouville/Toda cases. From the results of section three, we have
seen explicitly that in the important case of imaginary free-field momentum
which arises in non-critical string theories, the operator-state map does
not generally produce single interacting-theory eigenfunctions. The
ramifications of this observation need further study.

     Although Liouville and Toda gravities have already been studied in
considerable detail, the literature reveals different approaches to the
subject that have not been fully cross-fertilised. The above
considerations of the operator-state map might be cited as an example of this.
One would like to understand better how the background charges needed for
conformal-anomaly cancellation affect the canonical transformation to free
fields. A standard approach in much of the literature is to use the
operator-state map to implement a shift of the vertex-operator momenta as in
(2.3). This is unproblematic in free-field string theories, but, as we have
noted above, the operator-state map does not always generate single
interacting-theory eigenstates. Implementation of the
canonical-transformations to free fields in the Liouville or Toda theories
after first using the operator-state map to shift away the background charges
is not an obviously consistent procedure. On the other hand, the proper way
to incorporate directly the effects of the background charges into the
free-field canonical transformations is not clear.

     Another sense in which the results of the different approaches to
non-critical string theory need to be interrelated concerns the implications
of recent results on the spectra of low-dimensional non-critical strings
obtained {\it via} BRST analysis [26,27]. At the minisuperspace level
that has been our main concern in this paper, there are no surprises.
But when one includes the oscillator excitations, non-critical string
theories develop significant new features not generally taken into
account in Liouville or Toda field-theory discussions. In the analysis of
the BRST cohomology problem to determine the spectra of ordinary strings
with one or two scalars [26,27], or of $W_3$ strings with two scalars
[36,37], it turns out that the physical states do not occur only in the
ghost-vacuum sector. Indeed, it is necessary to incorporate the full set
of cohomologically-nontrivial states in order to see the Weyl-group
symmetry that has played a central r\^ole in the analysis of the present
paper.

     To make the r\^ole of ghost excitations in the excited states more clear,
consider the known free-field spectrum of the two-scalar $W_3$ string, which
corresponds, in the interacting theory, to the case of pure Toda $W_3$
gravity as considered in section four. The physical states are, by
definition, states that are annihilated by the BRST operator, but that are
not BRST trivial. Since we have only two scalar matter fields, and since
there are
two constraints (from the spin-2 current and the spin-3 current), it follows
that there are no continuous degrees of freedom ({\it i.e.}\ there are no
transverse spacetime dimensions), so all physical states have discrete
momenta. The physical states in this case can be subdivided into two
categories.  First, consider the physical states that involve no excitations
of the ghost fields; these are generally built by acting with vertex
operators and matter-field excitations on the ghost vacuum.  Since there are
no transverse dimensions in the two-scalar string, however, there cannot be
any such BRST non-trivial states involving matter-field excitations. Thus,
the only
physical states of this ``standard ghost structure'' are the tachyons, built
by acting with matter vertex operators $e^{\alpha_1\varphi_1 +\alpha_2
\varphi_2}$ on the ghost vacuum.  The momenta of these, and indeed of {\it
all} physical states in the two-scalar $W_3$ string are quantized and take
the form
$$
\alpha_1=\ft17 Q_1\, k_1,\qquad \qquad \alpha_2=\ft17 Q_2\, k_2, \eqno(5.1)
$$
where $k_1$ and $k_2$ are integers.  There are six tachyonic physical
states, with momenta given by (5.1) with $(k_1,k_2)= (-6,-6)$, $(-6,-8)$,
$(-8,-6)$,
$(-8,-8)$, $(-7,-5)$, $(-7,-9)$ [14,38].  These are mapped into one
another under the action of the six-element Weyl group of $A_2\equiv
SU(3)$.  The Weyl group acts by reflection in the momentum plane, and is
most appropriately described in terms of its action on the shifted momenta
$\hat\alpha_i=\alpha_i + Q_i$.  Correspondingly, we may
define the shifted integers $(\hat k_1, \hat k_2)=(k_1+7, k_2+7)$. The
Toda momenta that we have used earlier in this article are
related to the $\hat k_i$ by
$$\eqalign{
\im p_1&=-\ft12\hat k_1+\ft16\hat k_2\cr
\im p_2&=-\ft13\hat k_2.\cr}\eqno(5.2)
$$
The Weyl group is generated by the two transformations
$R$ and $M$ discussed in section four, which map $(\hat k_1,\hat k_2)$
according to
$$
\eqalignno{ R:&\qquad (\hat k_1,\hat k_2)\longrightarrow \big(-\ft12(\hat
k_1+\hat k_2), -\ft12(3\hat k_1-\hat k_2)\big)&(5.3)\cr
M:&\qquad (\hat k_1,\hat k_2)\longrightarrow \big(-\ft12(\hat k_1-\hat
k_2), -\ft12(3\hat k_1+\hat k_2)\big).&(5.4)\cr}
$$
The six elements of the Weyl group are given by 1, $R$, $M$, $RM$, $M^2$ and
$RM^2$.  One can easily verify that the six tachyon momenta given above are
mapped into one another under these
transformations.\footnote{$^{\dag}$}{\tenfoot In comparing the
transformations (5.3, 5.4) with our earlier forms (4.1.8, 4.1.9), it is
necessary to recall the difference between the free-field limit
of the interacting Toda variables $(p,q)$ and the free-field variables
$(\tilde p,\tilde q)$. As noted in section four, one needs to
apply a basis-change transformation, {\it e.g.}\ $(p_1,q_1)\mapsto(-\tilde
p_1,-\tilde q_1)$, in order to relate the two.}

     The above six tachyons are the only physical states without
ghost excitations in the two-scalar $W_3$ string.  Thus, as it stands, the
minisuperspace discussion of Toda quantum mechanics given in section four is
really only applicable as a tachyon-sector approximation to Toda field theory
in the context of pure $W_3$ gravity. The two-scalar $W_3$ string has
infinitely many more physical states, but all except the tachyons have
``non-standard'' ghost structure, involving excitations of the ghosts
as well as of the matter fields.  Thus, they fall outside the context of our
section-four discussion. Ultimately, one would like to be able to extend
the discussion of Toda field theory to include the full spectrum of physical
states in $W_3$ gravity, including those of non-standard ghost structure. To
our knowledge, the analogous issues have not yet been addressed in the
literature even for Liouville gravity.  In Liouville theory, the
majority of physical states for the pure gravity case ({\it i.e.}\ for the
one-scalar string) also have non-standard ghost structure.  Thus, one can
expect that a full treatment of Liouville or Toda field theory should include
the ghosts as well as the matter fields.

     The higher-level physical states of the free-field two-scalar $W_3$
string also fall into multiplets under the Weyl group of $A_2$.  The states
in a multiplet all have the same level number $\ell$, but, unlike the case
of the tachyons at level $\ell=0$, the various members of an $\ell\ne0$
multiplet have different ghost numbers.  Further details may be found in
[37], but the essential structure is as follows.  A necessary condition that
must be satisfied by any physical state is that it be described by an
operator with conformal dimension zero: this is the mass-shell condition.
For a state at level $\ell$, this implies that the shifted integers
describing the momentum satisfy
$$
3\hat k_1^2 + \hat k_2^2 =4(12\ell+1).\eqno(5.5)
$$
Remarkably, it turns out [37] that {\it any} solution of (5.5) for
integers $\hat k_1$, $\hat k_2$ and $\ell$ gives momenta corresponding to
some physical state of the theory.  Now, it is easy to see that the
mass-shell condition (5.5) is preserved by the Weyl group generated by
$R$ and $M$ as given in (5.3, 5.4).  Thus, it follows that the Weyl
group acts on the momentum of any physical state to give other
physical-state momenta.  Consequently, one can associate the physical states
together into multiplets by the action of the Weyl group.  In the case of
the tachyons at $\ell=0$, they all have the standard ghost structure, and
hence in particular have the same ghost number.  For the higher-level states,
which have non-standard ghost structure, the physical states with
Weyl-related momenta at a given level $\ell$ do not all have the same ghost
number.  As we have seen, the Weyl group plays an essential r\^ole in
relating the Toda and the free-field wavefunctions and operators. For the
full non-critical $W_3$ string theory, these observations underline the
importance of correctly taking the ghosts into account.

\bigskip
\bigskip
\centerline{\bf ACKNOWLEDGMENTS}
\bigskip
We would like to thank H. Lu, M. Olshanetsky, J. Schnittger and Y. Tanii for
helpful discussions. For hospitality during the course of the work, C.N.P.
would like to thank Chalmers University (Gothenburg),  Imperial College
(London) and SISSA (Trieste); K.S.S. would like to thank Chalmers University,
J.I.N.R. (Dubna), the Laboratoire de Physique Th\'eorique de l'Ecole Normale
Sup\'erieure (Paris), SISSA and Texas A\&M University.

\np
\noindent {\bf Appendix. Other Toda wavefunction representations}
\bigskip

     The integral representations of the Liouville and Toda wavefunctions
quoted in the review by Olshanetsky and Perelomov [28] can be
economically derived using canonical transformations. Two of these were
obtained in [29] by constructing power series solutions for the
wavefunctions and then by simply stating the associated integral
representations. A third, the result of Vinogradov and Takhtadjan [30]
for the 3-body Toda wavefunction, was obtained using methods proper to number
theory. It is useful to have explicit elementary derivations of these
integral representations.

     The key difference between these other representations and those
given in the body of this paper is that they are not obtained by
transforming all the way to a free theory, but rather by transforming to an
interacting theory for which one particular solution can be found by
inspection. The result is that one has an eigenvalue-dependent
transformation which produces a particular solution. This can be a
powerful method for constructing particular solutions to a theory [22].

     The result of Vinogradov and Takhtadjan [30] for the 3-body Toda
wavefunction is the most easily obtained and so we shall begin with it.  We
start from the 3-body Toda Hamiltonian
$$
H={1\over 2}(p_1^2 +p_2^2+p_3^2) + e^{q_1-q_2}+e^{q_2-q_3}.\eqno(A.1)
$$
This can be transformed into center-of-mass coordinates in the Chevalley
basis of $A_2$ by the transformation\footnote{$^{\ddag}$}{\tenfoot The
inverse of ($A$.2) is\
$\eqalign{ (q_1,q_2,q_3)\ &\mapsto\ (\ft23 q_1 -\ft13q_2 -
\ft13 q_3, \ft13 q_1 +\ft13 q_2 - \ft23 q_3, q_1+q_2+q_3) \cr (p_1,p_2,p_3)\
&\mapsto\ \big(p_1 - p_2,p_2 - p_3,\ft13 (p_1 + p_2 + p_3)\big).\cr}$}
$$
\matrix{ q_1&\mapsto &\ \ q_1 + \ft13 q_3,\hfill \cr q_2&\mapsto &-q_1 +q_2 +
\ft13 q_3,\cr q_3&\mapsto &-q_2 + \ft13 q_3,\hfill \cr}\qquad
\matrix{ p_1 &\mapsto &\ft23 p_1 +
\ft13 p_2 + p_3\cr p_2 &\mapsto &-\ft13 p_1 + \ft13 p_2 + p_3\cr p_3 &\mapsto
&-\ft13 p_1 -
\ft23 p_2 + p_3.\cr}\eqno(A.2)
$$
The resulting Hamiltonian is then of the form used in section four, but
including the center-of-mass variables,
$$
H^a={1\over 3}(p_1^2+p_2^2 +p_1 p_2) +{3\over 2} p_3^2 + e^{2q_1 -q_2} +
e^{2q_2 -q_1}.\eqno(A.3)
$$

     The following sequence of
transformations\footnote{$^{\S}$}{\tenfoot The inverse transformation
${\cal P}^{-1}_{{\rm\scriptscriptstyle VT1}}$ takes
$(q_1,q_2)\mapsto \big(\ft12(2q_1 -q_2),\ft12(2q_2-q_1)\big)$.} then
reduces the Hamiltonian to a form for which an eigenfunction can be found
by inspection.
\crampest
$$
\matrix{ [{\rm VT1}]&{\cal P}_{{\rm\scriptscriptstyle
VT1}}:&\rlap{$\left\{\matrix{q_1
\mapsto{2\over 3}(2q_1+q_2),
\cr q_2 \mapsto {2\over 3}(2q_2+q_1),\cr}\right.$}\hskip
4.75cm&\rlap{$\matrix{ p_1 \mapsto p_1 -{1\over 2}p_2\cr p_2 \mapsto
p_2-{1\over 2}p_1\cr}$}
\hskip 5.1cm\cr\strut\cr [{\rm VT2}]&\exp(-i\ln 2
(p_1+p_2)):&\rlap{$\left\{\matrix{q_1 \mapsto q_1-
\ln2,\cr q_2 \mapsto q_2- \ln2,\cr}\right.$}\hskip 4.75cm&\rlap{$\matrix{ p_1
\mapsto p_1\cr p_2 \mapsto p_2\cr}$}\hskip 5.1cm\cr\strut\cr [{\rm
VT3}]&\exp(-ik (q_1-q_2)):&\rlap{$\left\{\matrix{q_1\mapsto q_1,\cr  q_2
\mapsto q_2,\cr}\right.$}\hskip 4.75cm&\rlap{$\matrix{ p_1 \mapsto p_1+k\cr
p_2 \mapsto p_2-k\cr}$}
\hskip 5.1cm\cr\strut\cr [{\rm VT4}]&{\Gamma(-i{p_1+p_2\over 2}) \over
\Gamma(-i{p_1+3k\over 2})
\Gamma(-i{p_2-3k \over 2}) }:&\rlap{$\left\{\matrix{e^{2q_1} \mapsto
{p_1+3k\over p_1+p_2} e^{2q_1},\cr e^{2q_2} \mapsto {p_2-3k\over p_1+p_2}
e^{2q_2},\cr}\right.$}\hskip 4.75cm&\rlap{$\matrix{ p_1 \mapsto p_1\cr p_2
\mapsto p_2\cr}$}\hskip 5.1cm\cr\strut\cr}
$$
\uncramp  Suppressing the free-particle kinetic term $\ft32 p_3^2$ describing
the motion  of the center of mass, the Hamiltonian transforms as follows:
$$
\tabskip=0pt plus1fil
\halign to\displaywidth{\tabskip=0pt $\hfil#$&$\hfil#$&${}#\hfil$
\tabskip=0pt \cr &4 H^a&= {4\over 3}(p_1^2+p_2^2 +p_1 p_2) + 4e^{2q_1 -q_2} +
4e^{2q_2 -q_1}.\cr
[{\rm VT1}]&\qquad\qquad&\mapsto p_1^2+p_2^2-p_1 p_2 +
4e^{2q_1} +  4e^{2q_2}\cr
[{\rm VT2}]&\qquad\qquad&\mapsto p_1^2+p_2^2-
p_1 p_2 + e^{2q_1} +
e^{2q_2}\cr
[{\rm VT3}]&\qquad\qquad&\mapsto p_1^2+3kp_1+p_2^2-3kp_2- p_1
p_2 +  e^{2 q_1} +e^{2 q_2}+ 3 k^2 \cr
[{\rm VT4}]&\qquad\qquad&\mapsto
{1\over p_1+p_2}  \big((p_1+3k)(p_1^2 +e^{2q_1}) +(p_2-3k)(p_2^2+e^{2q_2})
\big) + 3 k^2.\cr }
$$
Inspection shows that this final separable Hamiltonian has
$K_\nu(e^{q_1})K_\nu(e^{q_2})$  as an eigenfunction, with eigenvalue
$\ft14(-\nu^2+3k^2)$. It is also an eigenfunction of the transformed $W$
operator invariant with eigenvalue $-\ft38k(k^2+\nu^2)$.

     Applying the intertwining operator, we obtain an eigenfunction of the
Toda Hamiltonian ($A$.3) (with zero center-of-mass momentum):
$$
\eqalign{
\psi_{k,\nu}(q_1,q_2)&=N_{k,\nu}{\cal P}^{-1}_{{\rm\scriptscriptstyle
VT1}}\exp\big(i(p_1+p_2)\ln 2\big)
\exp\big(ik(q_1-q_2)\big)\cr &\qquad\qquad {\Gamma(-i{p_1+3k\over 2})
\Gamma(-i{p_2-3k
 \over 2})\over \Gamma(-i{p_1+p_2\over 2}) }
K_\nu(e^{q_1})K_\nu(e^{q_2}).}\eqno(A.4)
$$
Recognizing the product of Gamma functions as a Beta function, and using  a
familiar integral representation for the Beta function, one finds the
result, in the Chevalley basis,
$$
\eqalign{
&\psi_{k,\nu}\cr
&=N_{k,\nu}\,{\cal P}^{-1}_{{\rm\scriptscriptstyle VT1}}e^{\im
k(q_1-q_2)}
\int_0^{\infty} dt\,t^{-1-\ft12\im(p_2-3k)}(1+t)^{\ft12\im(p_1+p_2)}
K_\nu(2e^{q_1})K_\nu(2e^{q_2})\cr
&=N_{k,\nu}\,e^{\ft32\im k(q_1-q_2)}\int_0^{\infty}
dt\, t^{-1+\ft32\im k} K_\nu(2\sqrt{(1+t)}e^{\ft12(2q_1-q_2)})
K_\nu(2\sqrt{(1+t^{-1})}e^{\ft12(2q_2-q_1)}).\cr}\eqno(A.5)
$$
Upon making an asymptotic expansion, one finds that this integral corresponds
to the solution  (4.2.10) with $k=-(k_1+k_2)/3,\
\nu=i(k_1-k_2)$. Transforming back into the coordinates of the original
Hamiltonian ($A$.1), one finds
$$
\eqalignno{
\Psi_{k,\nu}&=N_{k,\nu}\,\exp\Big(\ft12\im k(q_1-2q_2+q_3)\Big)\int_0^{\infty}
dt\,  t^{-1+\ft32\im k}\times&\cr
&\qquad K_\nu(2\sqrt{(1+t)}e^{\ft12(q_1-q_2)})
K_\nu(2\sqrt{(1+t^{-1})}e^{\ft12(q_2-q_3)}).&(A.6)\cr}
$$
With the identifications $k=-\ft12\im(t-s)$, $\nu=-1+\ft12(3s+t)$, this
agrees with the result of Vinogradov and Takhtadjan, as quoted in [28].

     The Liouville eigenfunctions can be found by a similar procedure.  The
motivating idea for this form is to factor out the asymptotic plane-wave
behaviour of the Liouville wavefunction and to express the solution in terms
of this plane wave.  The sequence of transformations and their effect  on
the Hamiltonian $H=p^2+e^q$ (note the difference in the argument of the
exponential from the Hamiltonian given in section two) is
$$
\openup1\jot \tabskip=0pt plus1fil \halign to\displaywidth{\tabskip=0pt
$\hfil#$&$\hfil#$&${}#\hfil$\tabskip=0pt plus1fil \cr
 &\  H=&\ p^2+e^q\cr
 e^{-ikq}:& \qquad\qquad&\mapsto (p+k)^2+e^q  \cr
 {\cal P}_{\ln q}:& \qquad\qquad&\mapsto (qp+k)+q  \cr
 {\cal I}:& \qquad\qquad&\mapsto (-pq+k)^2 +p  \cr
 {\cal P}_{1/q}:& \qquad\qquad&\mapsto (q^2 p q^{-1}+k)^2 -q^2 p  \cr
 q^{2ik-1}:& \qquad\qquad&\mapsto (qp-k)^2 -qpq+2kq=(qp-2k)q(p-1)+k^2. \cr}
$$
The final Hamiltonian has $e^{iq}$ as an eigenfunction, with eigenvalue
$k^2$.  The wavefunction it produces is
$$\eqalignno{
\psi_k(q)&=N_k\,e^{\im kq}\,{\cal P}_{e^q}\,{\cal I}^{-1}\,{\cal
P}^{-1}_{1/q}\,q^{1-2\im k}\,e^{\im q}&\cr
&=N_k\,e^{\im kq}\int_0^{\infty} du\,
u^{-1-2ik}e^{\im (u-e^qu^{-1})}.&(A.7)\cr}
$$
Rotating the contour of the $u$ integration onto the positive imaginary axis
and redefining $u\mapsto u e^{\im\pi/2}$, one finds the result
$$
\eqalignno{
\psi_k(q)&=N_k\,e^{\pi k}e^{\im kq}\int_0^{\infty} du\,
u^{-1-2\im k}e^{-(u+e^q/u)}&\cr &=2N_k\,e^{\pi k}K_{2\im k}(2
e^{\ft12q}).&(A.8)\cr}
$$

     The Toda solution given by Bruschi {\it et al.}\ [29] can similarly be
found by this approach. The key idea, as in the Liouville case, is to begin
by factoring out the asymptotic plane-wave behaviour.  We shall denote the
momentum of this asymptotic plane wave  by
$\lambda=(\lambda_1,\lambda_2,\lambda_3)$ in the coordinate  system where the
Toda Hamiltonian takes the form ($A$.1). We also let
$a=\lambda_1-\lambda_2,\ b=\lambda_2-\lambda_3$. The sequence of
transformations to a theory for which an eigenfunction can be recognized by
inspection is
\crampest
$$
\openup1\jot \tabskip=0pt plus1fil \halign to\displaywidth{\tabskip=0pt
$\hfil#$&$\hfil#$&${}#\hfil$\tabskip=0pt \cr &\  H=&\
\ft12(p_1^2+p_2^2+p_3^2)+e^{q_1-q_2} +e^{q_2-q_3}\cr
 e^{-\im\lambda\cdot q}:&\qquad&\mapsto {1\over 2}\big((p_1+\lambda_1)^2
 +(p_2+\lambda_2)^2+(p_3+\lambda_3)^2\big)+e^{q_1-q_2} +e^{q_2-q_3} \cr
 {\cal P}_{{\rm\scriptscriptstyle C.O.M.}}:&\qquad&\mapsto p_1^2+p_2^2-p_1
 p_2 + a p_1
 + b p_2 + e^{q_1} +e^{q_2} \cr &\qquad&\qquad +{3\over 2}p_3^2
+(\lambda_1+\lambda_2+\lambda_3)p_3  +{1\over
2}(\lambda_1^2+\lambda_2^2+\lambda_3^2)  \cr
 {\Gamma(1+\im(a+p_1)) \over \Gamma(-\im(b+p_2))
\Gamma(1+\im(a+b+p_1+p_2)) }:&\qquad& \mapsto {1\over a+b+p_1+p_2}
\bigl[(a+p_1)(p_1^2+(a+b) p_1 + e^{q_1}) \cr &\qquad&\qquad
+(b+p_2)(p_2^2+(a+b) p_2 -e^{q_2})\bigr]  \cr &\qquad&\qquad  +
{3\over 2}p_3^2 +(\lambda_1+\lambda_2+\lambda_3)p_3  +{1\over
2}(\lambda_1^2+\lambda_2^2+\lambda_3^2)  \cr
 e^{\ft12\im(a+b)(q_1+q_2)}:&\qquad& \mapsto {1\over
p_1+p_2}\bigl[(p_1+\ft12(a-b))(p_1^2+
 e^{q_1}) +(p_2+\ft12(b-a))(p_2^2 -e^{q_2})\bigr]  \cr &\qquad&\qquad
+{3\over 2}p_3^2  +(\lambda_1+\lambda_2+\lambda_3)p_3  -{(a+b)^2\over
4}+{1\over 2}(\lambda_1^2+\lambda_2^2+\lambda_3^2) \cr}
$$
\uncramp
where ${\cal P}_{{\rm\scriptscriptstyle C.O.M.}}:(q_1,q_2,q_3)\mapsto
\big(\ft13(2q_1+q_2+q_3),
\ft13(-q_1+q_2+q_3),{1\over 3}(-q_1-2q_2+q_3)\big)$. The final Hamiltonian
in this sequence has $K_\nu(2e^{q_1/2})H_\nu^{(1)}(2e^{q_2/2})$ as an
eigenfunction with eigenvalue $-\ft14\nu^2-\ft14(a+b)^2+\ft12(\lambda_1^2+
\lambda_2^2+\lambda_3^2)$. Choosing $\nu=\im(a+b)$ gives an eigenvalue
$\ft12(\lambda_1^2+\lambda_2^2+\lambda_3^2)$ for a state which is
asymptotically free, and with asymptotic momentum
$\lambda=(\lambda_1,\lambda_2,\lambda_3)$. This corresponds to the Toda
solution of Bruschi {\it et al.}  Assembling the intertwining operator, the
solution is
$$\eqalign{
\Psi_\lambda(q)= N_\lambda\,e^{\im\lambda\cdot q} {\cal
P}^{-1}_{{\rm\scriptscriptstyle C.O.M.}}&{\Gamma\big(-\im(b+p_2)\big)
\Gamma\big(1+\im(a+b+p_1+p_2)\big) \over \Gamma\big(1+\im(a+p_1)\big)}
e^{-\ft12\im(a+b)(q_1+q_2)}\cr
&K_{\im(a+b)}(2e^{\ft12q_1})H_{\im(a+b)}^{(1)}(2e^{\ft12q_2}).}\eqno(A.9)
$$
The product of Gamma functions is once again a Beta function and has the
integral representation (with parameters at the limits of convergence)
$$
{\Gamma\big(-\im(b+p_2)\big) \Gamma\big(1+\im(a+b+p_1+p_2)\big) \over
\Gamma\big
(1+\im(a+p_1)\big)}= e^{-(b+p_2)\pi}\int_0^\infty d\tilde z_3\, (1-\tilde
z_3)^{-\im a -1 -\im p_1} \tilde z_3^{-\im b -1 -\im p_2}.\eqno(A.10)
$$
Allowing the translation operators in the Beta-function integral to act,
this gives the integral representation
$$
\eqalignno{
\Psi_\lambda(q)={-2\im N_\lambda\over\pi}\,e^{\im\lambda\cdot q}
&e^{-b\pi}
\int_0^\infty d\tilde z_3\, (1-\tilde z_3)^{-\im a -1}
\tilde z_3^{-\im b -1}e^{-\ft12\im(a+b)(q_1-q_3)}&\cr
&K_{\im(a+b)}\biggr( {2\over\sqrt{1-\tilde z_3}}e^{{1\over 2}(q_1-q_2)}
\biggl)  K_{\im(a+b)}\biggr({2\over\sqrt{\tilde z_3}}e^{\ft12
(q_2-q_3)}\biggl)\Big).&(A.11)\cr}
$$
The integral representation implied by $(A.7, A.8)$ can be used for the
Bessel functions here to give
$$
\eqalign{
\Psi_\lambda(q)&={-\im N_\lambda\over2\pi}e^{-\pi(a+2b)}e^{\im\lambda\cdot q}
\int_0^\infty d\tilde z_3\, (1-\tilde z_3)^{-\im a -1}
\tilde z_3^{-\im b -1} \cr &\qquad \int_0^\infty dz_1 \int_0^\infty dz_2\,
(z_1  z_2)^{-\im(a+b)-1} e^{\im(z_1+z_2)}\exp\Big(-\im{e^{q_1-q_2}\over
z_2(1-\tilde z_3)} -\im {e^{q_2-q_3}\over z_1 \tilde z_3}\Big).\cr}
\eqno(A.12)
$$
Finally, absorbing the constant factors into the  normalization
and making the change of variables $\tilde z_3=z_3/(z_1 z_2)$, one reaches
the form of Bruschi {\it et al.}
$$
\eqalignno{
\Psi_\lambda(q)&=N'_\lambda\,e^{\im\lambda\cdot q}
\int\limits_0^\infty\int\limits_0^\infty\int\limits_0^\infty dz_1 dz_2 dz_3\,
(z_1
z_2-z_3)^{\im (\lambda_2-\lambda_1) -1}
z_3^{\im (\lambda_3-\lambda_2) -1}&\cr
&\qquad\qquad\qquad e^{\im(z_1+z_2)}\exp\Big(-\im{z_1 \over z_1 z_2-
z_3}e^{q_1-q_2} -\im {z_2\over  z_3} e^{q_2-q_3}\Big).&(A.13)}
$$
Note that this corrects some typos in the formula quoted in [28], most
importantly concerning the range of integration.  Note also that this is
an eigenfunction of ($A$.1) and not of the Hamiltonian of the form
discussed in Section 12 of [28].

\bigskip\bigskip
\singlespace
\centerline{\bf REFERENCES}
\frenchspacing
\bigskip

\item{[1]}A.M. Polyakov, {\sl Phys. Lett.} {\bf B103} (1981) 207, 211.

\item{[2]}E. D'Hoker and D.H. Phong, {\sl Rev. Mod. Phys.} {\bf 60} (1988)
917;\nl
E. D'Hoker, ``Lecture notes on 2-$D$ quantum gravity and Liouville theory,''

\item{[3]}L. Alvarez-Gaum\'e, ``Topics in Liouville theory,'' in the
proceedings of the Trieste Spring School, 1991 (World Scientific, 1992).

\item{[4]}N. Seiberg, ``Notes on quantum Liouville theory and quantum
gravity,'' in {\it Common trends in mathematics and quantum field theory},
Proc. of the 1990 Yukawa International Seminar, {\sl Prog. Theor. Phys.}
Suppl. 102 (1991).

\item{[5]}P. Ginsparg and G. Moore, ``Lectures on 2$D$ gravity and 2$D$
string theory,'' presented at the TASI Summer School, Boulder, 1992,
hep-th/9304011.

\item{[6]}D. Friedan, ``Introduction to Polyakov's string model,'' in {\it
Recent Advances in Field Theory and Statistical Physics,} eds J.-B. Zuber
and R. Stora, (World Scientific, Singapore, 1986);\nl
O. Alvarez, in {\it Unified String Theory,} eds M. Green and D.
Gross (World Scientific, 1986).

\item{[7]}J.-L. Gervais and A. Neveu, {\sl Nucl. Phys.} {\bf B199} (1982)
59; {\bf B209} (1982) 125; {\bf B224} (1983) 329; {\bf B238} (1984) 125;
{\bf B238} (1984) 396; {\bf B257 [FS14]} (1985) 59; {\bf B264} (1986) 557;
{\sl Commun. Math. Phys.} {\bf 100} (1985) 15; {\sl Phys. Lett.} {\bf B151}
(1985) 271.

\item{[8]}J.-L. Gervais, {\sl Commun. Math. Phys.} {\bf 130} (1990) 257;
{\bf 138} (1991) 301; {\sl Phys. Lett.} {\bf B243} (1990) 85; {\sl Nucl.
Phys.} {\bf B391} (1993) 287.

\item{[9]}J.-L. Gervais and J. Schnittger, ``The many faces of the
quantum Liouville exponentials,'' LPTENS-93/30, hep-th/9308134.

\item{[10]}H.J. Otto and G. Weigt, {\sl Phys. Lett.} {\bf B159} (1985)
341; {\sl Z. Phys.} {\bf C31} (1986) 219;\nl
G. Weigt, {\sl Phys. Lett.} {\bf B277} (1992) 79; ``Canonical
quantization of the Liouville theory, quantum group structures and
correlation functions,'' in {\it Pathways to fundamental theories}, Proc.
Johns Hopkins Workshop on Current Problems in Particle Theory 16 (World
Scientific, 1993).

\item{[11]}T.L. Curtright and C.B. Thorn, {\sl Phys. Rev. Lett.} {\bf 48}
(1982) 1309;\nl
E. Braaten, T.L. Curtright and C.B. Thorn, {\sl Phys. Lett.} {\bf B118}
(1982) 115; {\sl Ann. Phys.} {\bf 147} (1983) 365;\nl
E. Braaten, T.L. Curtright, G. Ghandour and C.B. Thorn, {\sl Phys. Rev.
Lett.} {\bf 51} (1983) 19; {\sl Ann. Phys.} {\bf 153} (1984) 147.

\item{[12]}Y. Kazama and H. Nicolai, ``On the exact operator formalism of
two-dimensional Liouville quantum gravity in Minkowski spacetime,'' DESY
93-043, UT Komaba 93-6, hep-th/9305023.

\item{[13]}A.B. Zamolodchikov, {\sl Teor. Mat. Fiz.} {\bf 65} (1985) 1205.

\item{[14]}S.R. Das, A. Dhar and S.K. Rama, {\sl Mod. Phys. Lett.}
{\bf A6} (1991) 3055; {\sl Int. J. Mod. Phys.} {\bf A7} (1992) 2295.

\item{[15]}J. Thierry-Mieg, {\sl Phys. Lett.} {\bf B197} (1987) 368.

\item{[16]}L.J. Romans, {\sl Nucl.  Phys.} {\bf B352} (1991) 829.

\item{[17]}M. Bershadsky, W. Lerche, D. Nemeschansky and N.P. Warner,
{\sl Phys. Lett.} {\bf B292} (1992) 35; {\sl Nucl. Phys.} {\bf B401} (1993)
304.

\item{[18]}M. Toda, {\sl Phys. Rep.} {\bf 18} (1975) 1.

\item{[19]}A. Anderson and R. Camporesi, {\sl Commun. Math. Phys.} {\bf
130} (1990) 61;\nl
A. Anderson, {\sl Phys. Lett.} {\bf B319} (1993) 157.

\item{[20]}A. Anderson, {\sl Phys. Rev.} {\bf D47} (1993) 4458.

\item{[21]}A. Anderson, ``Canonical transformations in quantum
mechanics,'' Imperial/TP/92-93/31, hep-th/9305054.

\item{[22]}A. Anderson, ``Special functions from quantum canonical
transformations,'' \nl
Imperial/TP/93-94/5, hep-th/9310168.

\item{[23]}P. Mansfield, {\sl Nucl. Phys.} {\bf B208} (1982) 277.

\item{[24]}D.I. Olive, ``Lectures on gauge theories and Lie algebras with
some applications to spontaneous symmetry breaking and integrable
dynamical systems,'' lectures given at University of Virginia,
Charlottesville, Imperial College preprint, 1982.

\item{[25]}E. D'Hoker and R. Jackiw, {\sl Phys. Rev.} {\bf D26} (1982)
3517;\nl
A. Kihlberg, {\sl Phys. Rev.} {\bf D27} (1983) 2542;\nl
R. Marnelius, {\sl Nucl. Phys.} {\bf B261} (1985) 319.

\item{[26]}B.H. Lian and G.J. Zuckerman, {\sl Phys. Lett.} {\bf 254B}
(1991) 417; {\sl Phys. Lett.} {\bf 266B} (1991) 21; {\sl Commun. Math.
Phys.} {\bf 145} (1992) 561.

\item{[27]}E. Witten, {\sl Nucl. Phys.} {\bf B373} (1992) 187;\nl
E. Witten and B. Zwiebach, {\sl Nucl. Phys.} {\bf B377} (1992) 55.

\item{[28]}M.A. Olshanetsky and A.M. Perelomov, {\sl Phys. Rep.} {\bf 94}
(1983) 313.

\item{[29]}M. Bruschi, D. Levi, M.A. Olshanetsky, A.M. Perelomov and O.
Ragnisco, {\sl Phys. Lett.} {\bf A88} (1982) 7.

\item{[30]}A.I. Vinogradov and A.A. Takhtadjan, ``Theory of the Eisenstein
series for the group $SL(3,\R)$ and its application to a binary problem,
I,'' {\sl Notes at the LOMI seminars} {\bf 76} (1978) 5.

\item{[31]}J. Polchinski, ``Remarks on the Liouville field theory,'' in
{\it Strings '90}, eds R. Arnowitt, R. Bryan, M.J. Duff, D. Nanopoulos,
C.N. Pope and E. Sezgin (World Scientific, 1991).

\item{[32]}L. Johansson and R. Marnelius, {\sl Nucl. Phys.} {\bf B254}
(1985) 201;\nl
C.R. Preitschopf and C.B. Thorn, {\sl Phys. Lett.} {\bf B250} (1990) 79.

\item{[33]}T. Curtright, ``Quantum B\"acklund transformations and
conformal algebras,'' in {\it Differential Geometric Methods in
Theoretical Physics}, eds L.L. Chau and W. Nahm (Plenum Press, NY, 1990).

\item{[34]}G.N. Watson, {\it A treatise on the theory of Bessel
functions}, 2'nd edition, (Cambridge University Press, 1944).

\item{[35]}H. Flaschka, {\sl Phys. Rev.} {\bf B9} (1974) 1924.

\item{[36]}S.K. Rama, {\sl Mod.\ Phys.\ Lett.}\ {\bf A6} (1991) 3531;\nl
C.N. Pope, E. Sezgin, K.S. Stelle and X.J. Wang, {\sl Phys. Lett.}
{\bf B299} (1993) 247;\nl
H. Lu, C.N. Pope, S. Schrans and X.J.
Wang, {\sl Nucl. Phys.} {\bf B408} (1993) 3;\nl
P. Bouwknegt, J. McCarthy and K. Pilch, {\sl Commun. Math. Phys.}
{\bf 145} (1992) 541.

\item{[37]}H. Lu, C.N. Pope, X.J. Wang and K.W. Xu,  ``The complete
cohomology of the $W_3$ string,'' preprint CTP TAMU-50/93, hep-th/9309041,
to appear in {\sl Class. Quantum Grav.}

\item{[38]}C.N. Pope, L.J. Romans and K.S. Stelle, {\sl Phys. lett.} {\bf
B268} (1991) 167; {\sl Phys. lett.} {\bf B269} (1991) 287;\nl
C.N. Pope, L.J. Romans, E. Sezgin and K.S. Stelle, {\sl Phys. Lett.}
{\bf 274B} (1992) 298;\nl
H. Lu, B.E.W. Nilsson, C.N. Pope, K.S. Stelle and P.C. West, {\sl Int. J.
Mod. Phys.} {\bf A8} (1993) 4071.

\bye